\documentclass[aps,prb,reprint,superscriptaddress,longbibliography]{revtex4-2}
\usepackage{graphicx}
\usepackage{textcomp}
\usepackage{gensymb}
\usepackage{amsmath, amssymb}
\usepackage[colorlinks = true, citecolor = magenta]{hyperref}
\usepackage{braket}
\usepackage{natbib}
\usepackage[utf8]{inputenc}

\begin{document}

\title{Entanglement Dynamics in a Two–Transmon Qubit System under Continuous Measurement and Postselection}
% \title{Tunable transverse and longitudinal couplings in a fluxonium-based electromechanical device}

\author{Roson~Nongthombam}
\email{n.roson@iitg.ac.in}
\affiliation{Department of Physics, Indian Institute of Technology Guwahati, Guwahati-781039, (India)}

\author{Amarendra~K.~Sarma}
\email{aksarma@iitg.ac.in}
\affiliation{Department of Physics, Indian Institute of Technology Guwahati, Guwahati-781039, (India)}

\date{\today}

\begin{abstract}
We investigate the role of continuous measurement and postselection in the dynamics and entanglement of a transmon-cavity-transmon coupled system. In the dispersive regime, characterized by a large detuning between the transmons and the cavity, the two transmons interact via virtual excitation of the cavity, giving rise to an effective transmon-transmon coupling. In addition to this coherent interaction, each transmon undergoes spontaneous emission, which is continuously monitored through independent detection channels. By incorporating realistic detector inefficiencies, we analyze both efficient and imperfect monitoring scenarios and demonstrate that postselection significantly slows down the decay of entanglement compared to the unmonitored case.
We formulate the stochastic master equation for the coupled system, derive the corresponding postselected master equation, and investigate the dynamics through the Liouvillian superoperator spectrum. In the interaction frame, we identify the emergence of an exceptional point and characterize the associated broken and unbroken $\mathcal{PT}$-symmetric phases. We show how these phases influence the system dynamics and the corresponding entanglement behavior.
Our results provide insight into how continuous measurement and postselection affect entanglement in dissipative quantum systems, with potential applications in quantum information processing.

\end{abstract}

\maketitle
%%%%%%%%%%%%%%%%%%%%%
\section{Introduction}
A quantum system interacting with its environment continuously becomes entangled with it. 
The combined system--environment forms a closed system and evolves unitarily, leading to entanglement between the system and the environment. 
Upon measuring the state of the environment, the system collapses into a corresponding conditional state.
If the environment is continuously monitored, and since the environment continuously interacts with the system, the conditional state of the system evolves purely, giving rise to a single quantum trajectory. However, if the measurement outcomes of the environment are not accessible, the state of the system is known by tracing out the environmental degrees of freedom. 
This corresponds to taking an ensemble average over many quantum trajectories, resulting in a mixed state of the system. 

Measuring the state of the environment and extracting the quantum trajectories of a system has been studied extensively, both experimentally and theoretically. 
This includes probing the location of a single electron in a double quantum dot using a capacitively coupled quantum point contact \cite{Ashhab_2009,Korotkov}, as well as probing the state of a superconducting transmon qubit dispersively coupled to a three-dimensional microwave cavity \cite{naghiloo2016,PhysRevX.6.011002,weber2014,naghiloo2019}.
Quantum trajectory unraveling can be performed using various measurement schemes, including photodetection, where the measurement outcomes correspond to $0$ and $1$ in the Fock basis \cite{wiseman2009,steck2007quantum,lewalle2020,PhysRevResearch.6.L032057,mlmer1993}, and homodyne or heterodyne detection, which measure the quadratures of the environmental field \cite{lewallethesis,jordan2016,PhysRevA.96.053807,PhysRevX.6.011002,PhysRevA.96.022104,hmwiseman1996,PhysRevLett.70.548,weber2014,PhysRevA.89.023827,etde_21197676,PhysRevA.34.1642,PhysRevLett.52.1657,mlmer1993}.
Furthermore, quantum trajectories can be broadly classified into two categories. 
In the first, only the state of the system is monitored. For example, a transmon qubit dispersively coupled to a cavity can be probed by measuring the phase shift of the cavity field to infer the state of the system \cite{PhysRevA.77.012112,murch2013,PhysRevA.94.042326,jacobs2006,hmwiseman1996}. In the second, the spontaneously emitted photons from the system are directly measured \cite{lewallethesis,PhysRevResearch.6.L032057,naghiloo2016,PhysRevX.6.011002,PhysRevA.96.053807,jordan2016,PhysRevA.89.023827,minev2019,ficheux2018}. In this case, one can postselect trajectories corresponding to spontaneous emission events, thereby effectively generating a non-Hermitian system that can exhibit $\mathcal{PT}$-symmetric phases \cite{naghiloo2019,Chen2021,Ming2019,PhysRevLett.128.110402,PhysRevA.101.062112,Cbender_1999,Bender_2007}.

In this work, we consider continuous measurement of a transmon-cavity-transmon coupled system. The system consists of two transmons coupled via a microwave cavity and immersed in a cold bath at a temperature of approximately $15~\mathrm{mK}$. At this temperature, both the transmons and the cavity remain in their ground states. By detuning the transmons from the cavity, an effective transmon-transmon interaction is generated via virtual excitation of the cavity.
Such effective coupling mediated by virtual cavity excitation has been extensively explored and is widely used in implementing two-qubit quantum gates for quantum computation and information processing \cite{PhysRevA.75.032329,blais2020,majer2007,sillanp2007,dicarlo2009,8633444,PhysRevA.83.063827,bialczak2010,PhysRevLett.108.057002,Arjan,PhysRevA.69.062320}. 
It is also employed in protocols such as entanglement swapping \cite{PhysRevLett.123.060502}.
In addition to the coherent transmon-transmon interaction, which governs the unitary evolution, each transmon undergoes spontaneous emission. 
Here, we consider continuous monitoring of the spontaneously emitted photons from each transmon and unravel the corresponding quantum trajectories. 
By postselecting trajectories associated with no quantum jump events \cite{naghiloo2019,Chen2021,PhysRevA.101.062112}, i.e., spontaneous emission processes, we study the average dynamics of these postselected trajectories and the resulting entanglement.
In realistic scenarios, monitoring is inherently inefficient due to losses in the transmission channel and detector inefficiencies. 
We incorporate this inefficiency into the dynamics. 
We find that, for both efficient and inefficient monitoring, the entanglement of the postselected average decays at a significantly slower rate compared to the unmonitored case.

Furthermore, we formulate the stochastic master equation for the transmon-transmon coupled system and derive the corresponding postselected master equation. We compare the results obtained from this master equation with those from the postselected trajectory ensemble average for consistency. 
We also analyze the dynamics using the Liouvillian superoperator spectrum.
In the interaction frame, we investigate the emergence of an exceptional point and the corresponding broken and unbroken $\mathcal{PT}$-symmetric phases, along with their associated entanglement properties.
Our study demonstrates the effect of continuous measurement on two coupled transmons undergoing spontaneous emission and highlights the role of postselection in the resulting entanglement between the transmons. These results provide insight into how entanglement can be preserved through postselected continuous measurement.

The paper is organized as follows. In Sec.~II, we describe how the effective transmon-transmon coupling is obtained and discuss the entanglement generated from this interaction. In Sec.~III, we present the measurement scheme for the transmons and obtain the quantum trajectories using the Kraus operator state-update rule. We then compare the entanglement obtained in the unmonitored case with that from perfect and imperfect efficient postselected average dynamics. In Sec.~IV, we analyze the dynamics using the Liouvillian superoperator and study the emergence of the exceptional point (EP) and the associated $\mathcal{PT}$-symmetric phases.

\begin{figure}
\centering
\includegraphics[width=85mm]{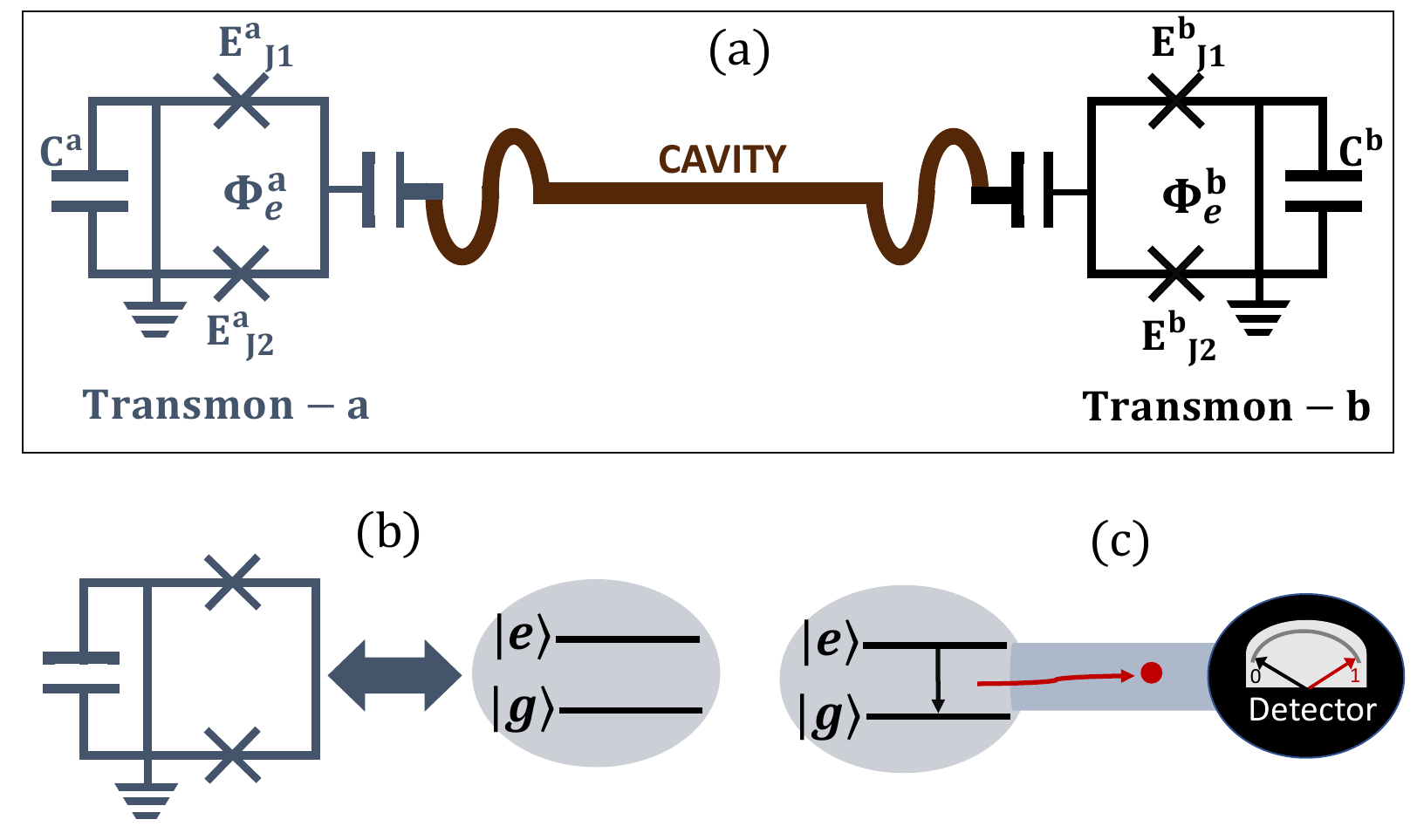}
\caption{(a) Schematic of a transmon--cavity--transmon coupled system, where two transmons (a and b) are capacitively coupled via a cavity. 
(b) Representation of the transmon qubit as an effective two-level system with ground and excited states $|g\rangle$ and $|e\rangle$. 
(c) Schematic of the monitoring of spontaneous emission from the transmon qubit. The emitted photon is collected through a transmission line and routed to a detector that registers its presence.}
\label{fig:System setup}
\end{figure}
%%%%%%%%%%%%%%%%%%%%%%%%%%%%%%%%%%%%%%%%%%%%%%%%%%%%%%%
\section{Generation of Entanglement in Two Transmons}
\begin{figure}
\centering
\includegraphics[width=35mm]{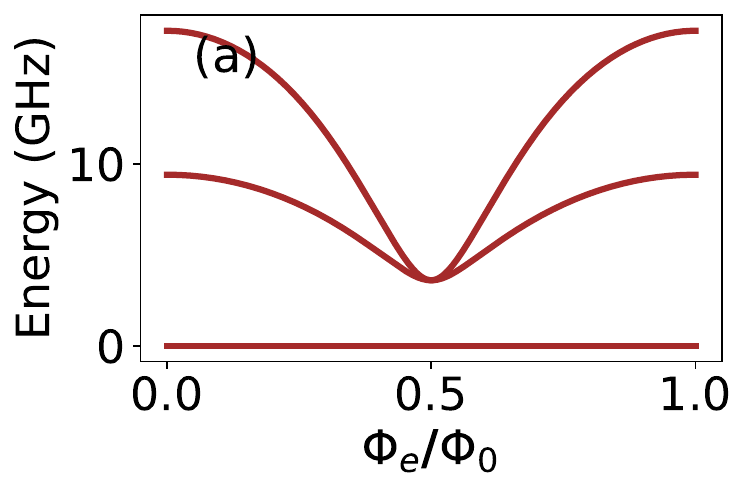}
\includegraphics[width=35mm]{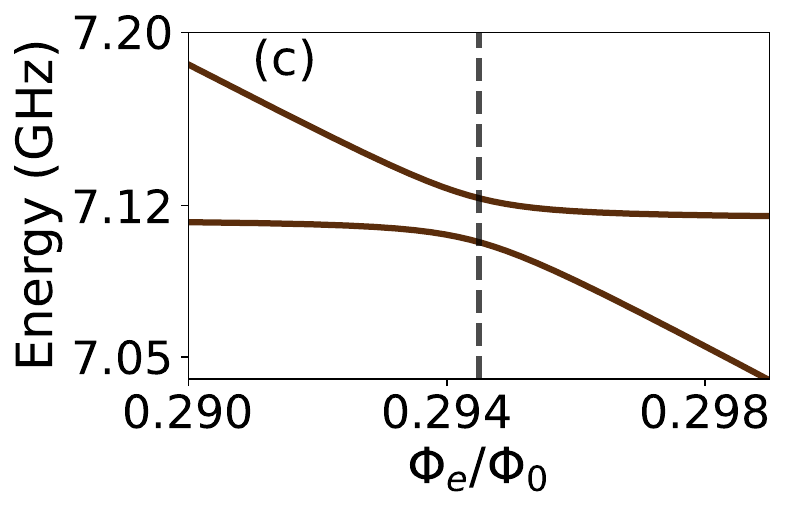}
\includegraphics[width=70mm]{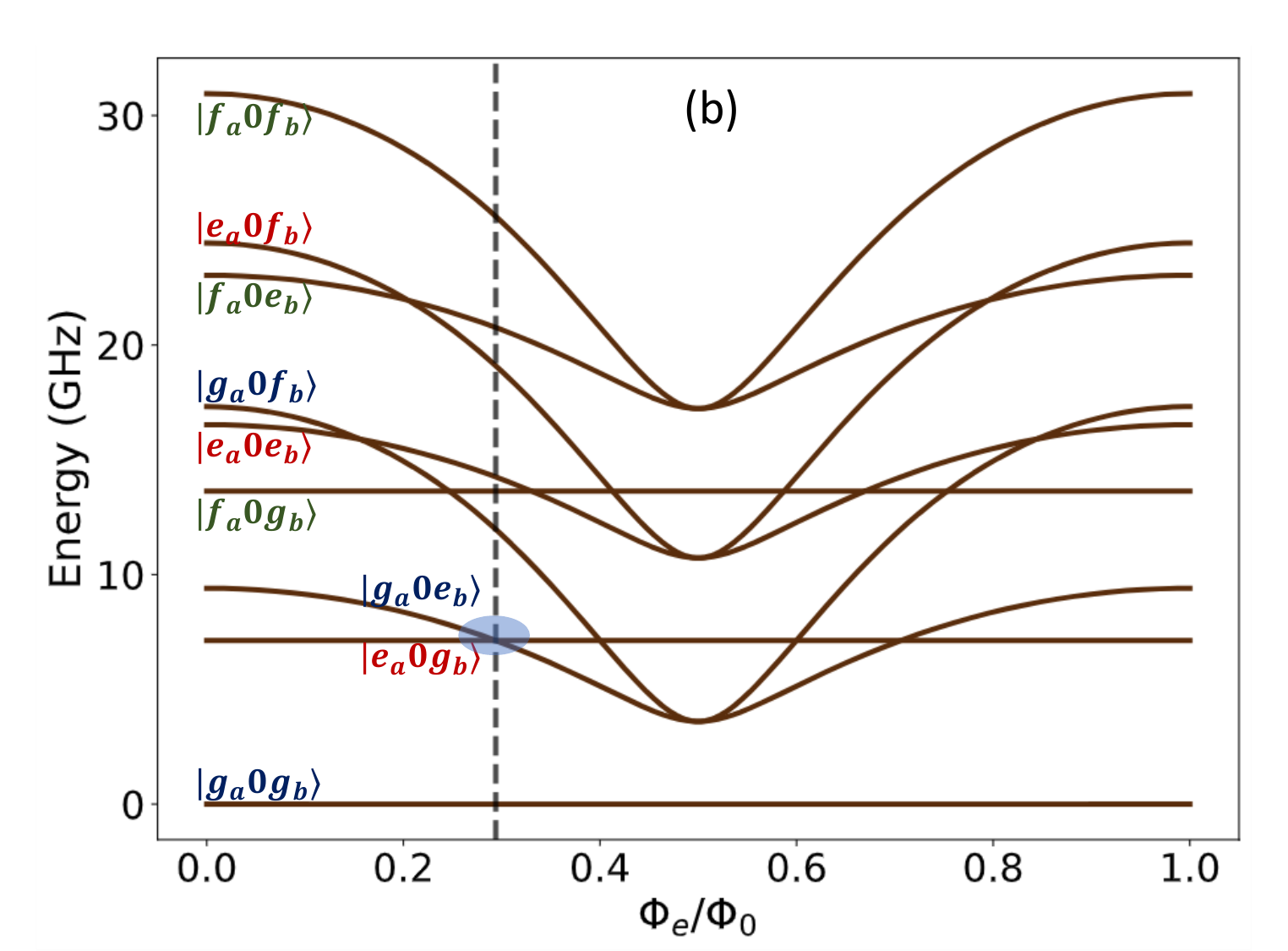}
\caption{Energy (eigenvalue) spectrum of (a) a single transmon and (b) a coupled transmon-cavity-transmon system. The spectrum is obtained by varying the externally applied flux $\Phi_e$. In panel (b), one of the transmons (denoted by $a$) is fixed at $\Phi_e = 0.3\,\Phi_0$, while the flux of the other transmon (denoted by $b$) is varied. For clarity, only the zero-excitation cavity manifold is shown. 
At the flux point $\Phi_e = 0.2945\,\Phi_0$ (indicated by the dotted line), the states $|e_a~0~g_b\rangle$ and $|g_a~0~e_b\rangle$ exhibit an avoided crossing. A zoomed-in view of this region, highlighted by the shaded circle in (b), is shown in (c). The energy gap at the avoided crossing is given by $2G_{eg}$ (refer Eq.\eqref{Eq:H_D_TT}).
The parameters used are as follows: for transmon $b$, $E_{J,\Sigma} = 15~\mathrm{GHz}$ and $E_C = 0.9~\mathrm{GHz}$; for transmon $a$, $E_{J,\Sigma} = 25~\mathrm{GHz}$ and $E_C = 0.5~\mathrm{GHz}$. The coupling strengths are $\zeta^a = \zeta^b = 0.3~\mathrm{GHz}$, and the cavity frequency is $\omega_c = 15~\mathrm{GHz}$.}
\label{fig:transmon_spec}
\end{figure}

%%%%%%%%%%%%%%%%%%%%%%%%%%%%%%%%%%%%%%%%%%%%%%%%%%%%%%
%
Consider two flux-tunable transmons connected via a microwave cavity through capacitive coupling as shown in Fig.~\ref{fig:System setup}. Each transmon is described by the Hamiltonian
\begin{equation}
\label{Eq:Transmon}
\hat{H} =
4E_C\, \hat{n}^2
-
E_{J,\Sigma}
\cos\!\left(\frac{\pi \Phi_{\mathrm{ext}}}{\Phi_0}\right)
\cos\hat{\phi}.
\end{equation}
Here $E_{J,\Sigma} = E_{J1} + E_{J2}$, where $E_{J1}$ and $E_{J2}$ are the Josephson energies of the two junctions. 
$\hat{n}$ is the Cooper-pair number operator, $\hat{\phi}$ is the superconducting phase operator
(with $[\hat{\phi},\hat{n}] = i$), $E_C = \frac{e^2}{2C}$ is the charging energy with $C$ the total capacitance,
$\Phi_{\mathrm{ext}}$ is the externally applied magnetic flux, and $\Phi_0 = \frac{h}{2e}$ is the superconducting flux quantum.
The eigenstates and eigenenergies are obtained by diagonalizing the transmon Hamiltonian given in Eq.~\eqref{Eq:Transmon}.
The energy spectrum of the first three levels of the transmon is shown in Fig.~\ref{fig:transmon_spec}(a).
The coupled transmon–cavity–transmon system is described by the Hamiltonian
\begin{align}
\label{Eq:Transmon-cavity-transmon}
\hat{H}_{\mathrm{TCT}} &=
\sum_{x = a,b} \hbar ~\zeta^x\, n^x_{g e}
\left(
|g_x\rangle\langle e_x|\, \hat{a}^\dagger
+
|e_x\rangle\langle g_x|\, \hat{a}
\right) 
\nonumber \\ &\quad 
+\sum_{x = a,b} \hbar ~\zeta^x\, n^x_{e f}
\left(
|e_x\rangle\langle f_x|\, \hat{a}^\dagger
+
|f_x\rangle\langle e_x|\, \hat{a}
\right) 
\nonumber \\ &\quad
+\sum_{\substack{x=a,b \\ i \in \{g,e,f\}}}
\hbar ~E^x_{i}\, |i_x\rangle\langle i_x| + \hbar\omega_c\hat{a}^\dagger\hat{a}.
\end{align}
Here, $a$ and $b$ denote the two transmons. $n^x_{g e} = \langle g_x | \hat{n}^x | e_x \rangle$ and $n^x_{e f} = \langle e_x | \hat{n}^x | f_x \rangle$ are the matrix element of the number operator ($\hat{n}^x$) for transmon $x$.
The quantities $E^x_{i}$ are the eigenenergies (in $Hz$) corresponding to the eigenstates $|i_x\rangle$ of transmon $x \in \{a,b\}$. 
Here, $\hat{a}$ ($\hat{a}^\dagger$) is the annihilation (creation) operator of the cavity mode with frequency $\omega_c$, and $\zeta^x$  are the coupling strengths between the cavity and the transmon $x$.
The energy spectrum of the coupled system is shown in Fig.~\ref{fig:transmon_spec}(b). 
The spectrum is obtained by keeping one of the transmons at a fixed external flux while varying the flux in the other transmon. 
For simplicity, only the zero-excitation manifold of the cavity is shown.
As seen in the figure, at a particular flux bias point the energies of the states $|e_a\,0\,g_b\rangle$ and $|g_a\,0\,e_b\rangle$ become degenerate. 
Due to the effective interaction between the two transmons mediated by the cavity, an avoided crossing appears between these states. 
This is highlighted in Fig.~\ref{fig:transmon_spec}(c).
Therefore, by tuning the external flux of the transmons to the avoided crossing point, one can observe the exchange interaction and generate entanglement between the two transmons.

The effective interaction between the two transmons can be evaluated by performing a dispersive transformation on the Hamiltonian in Eq.~\eqref{Eq:Transmon-cavity-transmon}. 
To achieve this, we detune the transmons from the cavity such that
\(
\eta_i^x =
\frac{\zeta^x\, n_{i,i+1}^x}{\left| E^x_{i+1} - E^x_i - \omega_c \right|} \ll 1.
\)
This condition defines the dispersive regime. 
The dispersive transformation eliminates the interaction term in Eq.~\eqref{Eq:Transmon-cavity-transmon} up to lowest order in $\eta_i^x$, yielding an effective transmon-transmon interaction. 
The transformed Hamiltonian is given by
\(
\hat{H}^{D}_{\mathrm{TCT}} = \hat{D}\,\hat{H}_{\mathrm{TCT}}\,\hat{D}^\dagger,
\)
where the unitary operator is defined as
\(
\hat{D} = \exp\!\left(\hat{S} - \hat{S}^\dagger\right),
\)
with
\begin{equation}
\hat{S} =
\sum_{\substack{x=a,b \\ i \in \{0,1,2\}}}
\beta_i^x \,\hat{a}\, |i+1\rangle_x \langle i|,
\end{equation}
% \vspace{-1.5mm}
and
% \vspace{-2mm}
\begin{equation}
\beta_i^x =
\frac{\lambda^x_{i,i+1}}{E^x_{i+1} - E^x_i - \omega_c}~~,
\quad
\lambda^x_{i,i+1}=\zeta^x~n^x_{i,i+1}~.
\end{equation}
Here, we identify the indices $i = 0,1,2$ with the transmon states $|g\rangle, |e\rangle, |f\rangle$, respectively. 
The transformed Hamiltonian upto the second order in $\lambda^x_{i,i+1}$ is given by
\begin{equation}
    \label{Eq:H_D_TCT}
    \hat{H}^D_{TCT}=\hat{H}_0 +\hat{H}_{LS}+\hat{H}_{AC}+\hat{H}_{C},
\end{equation}
% \onecolumngrid
%
% \begin{widetext}
where
\begin{equation}
\label{Eq:H_D_TCT_component}
\hat{H}_0
=
\sum_{\substack{x=a,b \\ i \in \{0,1,2\}}}
E^x_{i}~ |i\rangle_x\langle i| 
+
\omega_c\hat{a}^\dagger\hat{a},
\end{equation}
\vspace{-4mm}
\begin{equation}
\hat{H}_{AC}=\sum_{\substack{x=a,b \\ i \in \{0,1,2\}}}\lambda^x_{i,i+1}~\beta^x_i~\hat{a}^\dagger\hat{a}~\left(|i+1\rangle_x\langle i+1|-|i\rangle_x\langle i|\right),
\end{equation}
\vspace{-3mm}
\begin{equation}
\hat{H}_{LS}=+
\sum_{\substack{x=a,b \\ i \in \{0,1\}}}
\beta^x_{i} \lambda^x_{i,i+1}
|i+1\rangle_x\langle i+1|, 
\end{equation}
\vspace{-5mm}
\begin{align}
\hat{H}_{C}&=
\frac{1}{2}
\sum_{i=0}^{1}\sum_{k=0}^{1}
(\lambda^a_{i,i+1} \beta^b_{k} + \lambda^b_{k,k+1} \beta^a_{i})
\nonumber\\ & \quad
\left(
|i,k+1\rangle_{ab}\langle i+1,k|
+
|i+1,k\rangle_{ab}\langle i,k+1|
\right).
\end{align}
% \end{widetext}
% \twocolumngrid
%
Here, we set $\hbar = 1$. As can be seen, there is no direct exchange of excitation between the transmons and the cavity. 
$\hat{H}_{LS}$ and $\hat{H}_{AC}$ correspond to the Lamb shift and the ac Stark shift, respectively. 
The effective interaction between the two transmons is given by $\hat{H}_C$. 
If we assume the cavity to be initially in the ground state (which is typically the case in cQED experiments), the cavity components in the Eq.~\eqref{Eq:H_D_TCT} can be neglected yielding an effective two coupled transmons Hamiltonian.

Both the transmons and the cavity can undergo spontaneous emission by releasing photons into their respective baths. 
This process is described by the master equation
\begin{align}
\label{Eq:dissipative dynamics}
\frac{d\overline{\hat{\rho}}_{\mathrm{tct}}}{dt}
&=
-i\,[\hat{H}_{\mathrm{TCT}}, \overline{\hat{\rho}}_{\mathrm{tct}}]
+\sum_{x=a,b}
\Bigg[
\Gamma_{x,10}\,
\mathcal{D}[|0\rangle_x\langle 1|]\,
\overline{\hat{\rho}}_{\mathrm{tct}}
+
\nonumber\\[6pt]
&\quad
\Gamma_{x,21}\,
\mathcal{D}[|1\rangle_x\langle 2|]\,
\overline{\hat{\rho}}_{\mathrm{tct}}
\Bigg]  + \kappa\, \mathcal{D}[\hat{a}]\, \overline{\hat{\rho}}_{\mathrm{tct}}.
\end{align}
Here, $\overline{\hat{\rho}}_{\mathrm{tct}}$ is the density matrix of the transmon–cavity–transmon system, and $\hat{H}_{\mathrm{TCT}}$ is the system Hamiltonian defined in Eq.~\eqref{Eq:Transmon-cavity-transmon}. 
The parameter $\kappa$ denotes the decay rate of the cavity mode, while $\Gamma_{x,10}$ and $\Gamma_{x,21}$ represent the relaxation rates of transmon $x \in \{a,b\}$ for the transitions $|e\rangle \to |g\rangle$ and $|f\rangle \to |e\rangle$, respectively. 
The superoperator $\mathcal{D}[\hat{O}]$ is the Lindblad dissipator, defined as
\(
\mathcal{D}[\hat{O}]\,\overline{\hat{\rho}}
=
\hat{O}\overline{\hat{\rho}}\hat{O}^\dagger
-
\frac{1}{2}
\left\{
\hat{O}^\dagger \hat{O}, \overline{\hat{\rho}}
\right\},
\) where $\{\cdot,\cdot\}$ denotes the anticommutator.
In a typical experiment, the system is maintained at about $15~\mathrm{mK}$. The transmons and the cavity therefore remain in the ground state, and thermally induced excitations are neglected in the master equation.

\begin{figure*}[t]
\centering
\includegraphics[width=130mm]{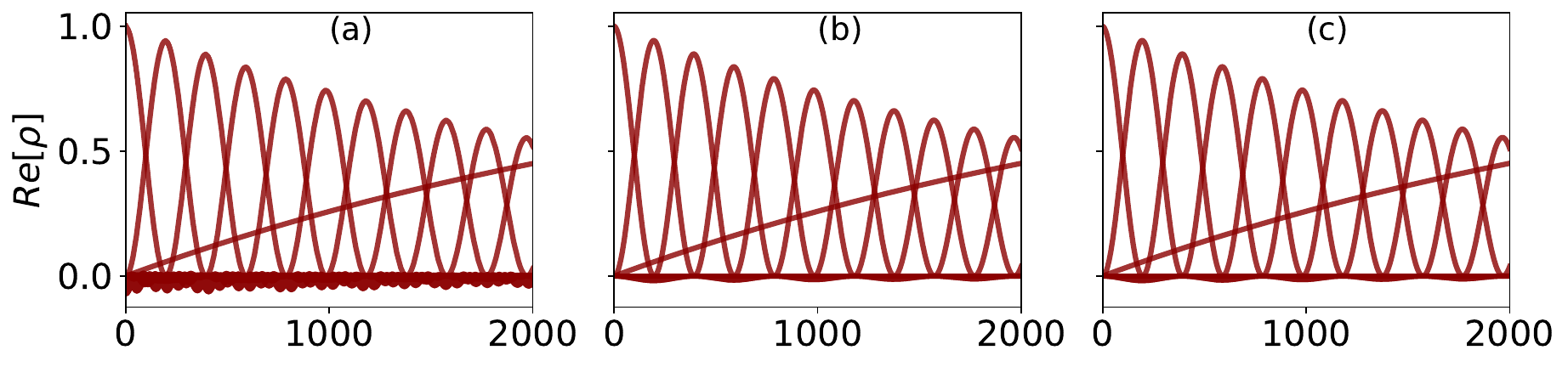}
\includegraphics[width=130mm]{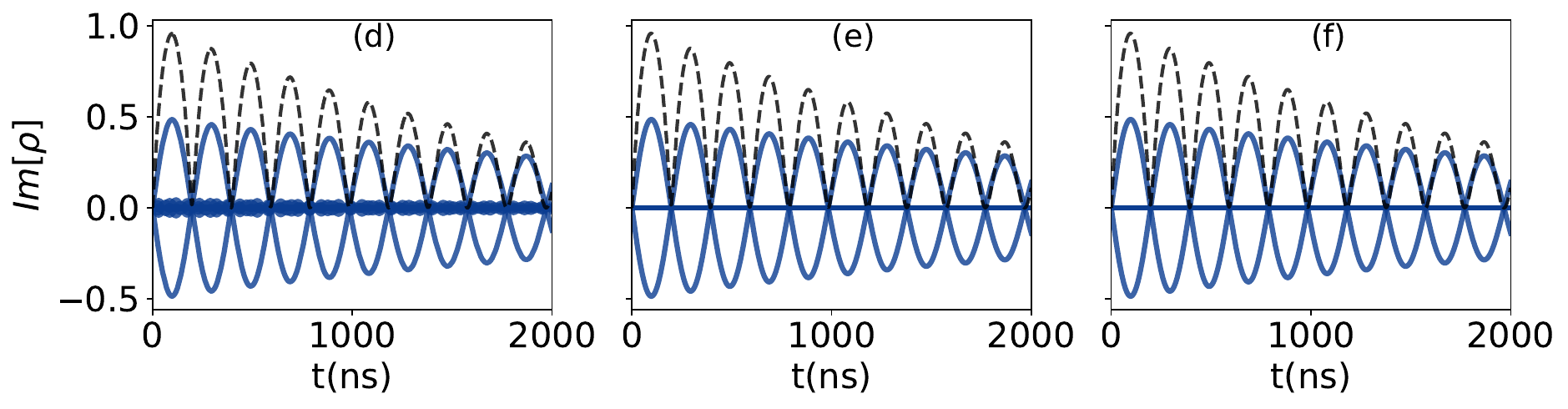}
\caption{Comparison of the real and imaginary parts of the density matrix elements for (a,d) the transmon-cavity-transmon coupled system [Eq.~\eqref{Eq:Transmon-cavity-transmon}], (b,e) the dispersive transmon-cavity-transmon system [Eq.~\eqref{Eq:H_D_TCT}], and (c,f) the effective transmon-transmon system [Eq.~\eqref{Eq:H_D_TT}]. 
The oscillations observed in (a,d) and (b,e) occur between the populations $\langle e_a0g_b | \overline{\hat{\rho}}_{\mathrm{tct}} | e_a0g_b \rangle$ and $\langle g_a0e_b | \overline{\hat{\rho}}_{\mathrm{tct}} | g_a0e_b \rangle$. These oscillations decay over time due to environmental dissipation. As a result the ground-state population $\langle g_a0g_b | \overline{\hat{\rho}}_{\mathrm{tct}} | g_a0g_b \rangle$ increases.
Since the cavity remains in its ground state throughout the evolution, the effective transmon-transmon model reproduces the same dynamics. In this case, the oscillations occur between $\langle e_ag_b | \overline{\hat{\rho}}_{\mathrm{tt}} | e_ag_b \rangle$ and $\langle g_ae_b | \overline{\hat{\rho}}_{\mathrm{tt}} | g_ae_b \rangle$, which also decay to the ground state $\langle g_ag_b | \overline{\hat{\rho}}_{\mathrm{tt}} | g_ag_b \rangle$ as is shown in (c,f).
In (d)-(f), the corresponding entanglement (defined in the text) between the two transmons is shown by the grey dashed curve, exhibiting identical behavior in all three cases.
This confirms that, in the absence of cavity excitation, the effective transmon-transmon model fully captures both the population dynamics and entanglement evolution of the full system.
The Lamb shift is adjusted into the bare frequencies in the numerical simulations.
The parameters used are as follows: the transmon and cavity parameters are the same as in Fig.~\ref{fig:transmon_spec}. The decay rates are $\Gamma_{a,10} = \Gamma_{b,10} = 0.3~\mathrm{MHz}$, $\Gamma_{a,21} = \Gamma_{b,21} = 0.2~\mathrm{MHz}$, and the cavity decay rate is $\kappa = 0.3~\mathrm{MHz}$.}
\label{fig:fig:density evolution with and without cav}
\end{figure*}

Choosing the external flux bias at the avoided crossing shown in Fig.~\ref{fig:transmon_spec} and detuning the two transmons from the cavity, the evolution of the density matrix elements of $\overline{\hat{\rho}}_{\mathrm{tct}}$ is shown in Fig.~\ref{fig:fig:density evolution with and without cav}(a,b). 
 The system is initially prepared in the state $|e_a 0 g_b\rangle$.
 As shown in the figure, oscillations in the populations $\langle e_a 0 g_b | \overline{\hat{\rho}}_{\mathrm{tct}} | e_a 0 g_b \rangle$ and $\langle g_a 0 e_b | \overline{\hat{\rho}}_{\mathrm{tct}} | g_a 0 e_b \rangle$ are observed, revealing coherent population exchange between the states $|e_a 0 g_b\rangle$ and $|g_a 0 e_b\rangle$. Due to spontaneous emission, this exchange gradually decays, and the system eventually relaxes to the ground state $|g_a0 g_b\rangle$, as seen in the figure.
Small oscillations involving higher cavity excitation are present but remain negligible and can be safely ignored. The coherence dynamics is shown in Fig.~\ref{fig:fig:density evolution with and without cav}(b), which involves the imaginary part of the density matrix element $\langle e_a 0 g_b | \overline{\hat{\rho}}_{\mathrm{tct}} | g_a 0 e_b \rangle$. 
The oscillations in both the populations and the coherences between the $|e\rangle$ and $|g\rangle$ states of the two transmons lead to the generation of entanglement between them.

Instead of evolving the Hamiltonian in Eq.~\eqref{Eq:Transmon-cavity-transmon} directly under the detuning condition, the dissipative dynamics can also be obtained by substituting the dispersive Hamiltonian in Eq.~\eqref{Eq:H_D_TCT} into the master equation in Eq.~\eqref{Eq:dissipative dynamics}. 
A similar behavior in both the population and coherence dynamics is observed, as shown in Fig.~\ref{fig:fig:density evolution with and without cav}(c,d). 
In evaluating the master equation, it is assumed that the cavity remains in the ground state, and the contribution from Purcell decay \cite{PhysRevA.79.013819,PhysRevA.77.060305,purcell1946} is negligible due to the large detuning. 
The validity of this approximation is evident from the close agreement between the dynamics generated by $\hat{H}_{\mathrm{TCT}}$ and $\hat{H}_{\mathrm{TCT}}^{D}$. 

Furthermore, the dissipative population and coherence dynamics can also be obtained using the dispersive Hamiltonian $\hat{H}_{\mathrm{TCT}}^{D}$ with the cavity degrees of freedom removed, since the cavity remains in the ground state \cite{PhysRevLett.123.060502}. 
This is shown in Fig.~\ref{fig:fig:density evolution with and without cav}(e,f), which demonstrates good agreement with the dynamics obtained from the full cavity-included dispersive model. 
In this case, the dissipative dynamics is governed by
\begin{align}
\label{Eq:dissipative dynamics wc}
\frac{d\overline{\hat{\rho}}_{\mathrm{tt}}}{dt}
&=
-i\,[\hat{H}^D_{\mathrm{TT}}, \overline{\hat{\rho}}_{\mathrm{tt}}]
+\sum_{x=a,b}
\Gamma_{x,10}\,
\mathcal{D}[|0\rangle_x\langle 1|]\,
\overline{\hat{\rho}}_{\mathrm{tt}},  
\end{align}
where 
\begin{align}
\label{Eq:H_D_TT}
\hat{H}^D_{\mathrm{TT}}
&=
\sum_{\substack{x=a,b \\ i \in \{0,1\}}}
E^x_{i} ~|i\rangle_x\langle i| 
+
\sum_{x=a,b }
\beta^x_{0} \lambda^x_{0,1}
|1\rangle_x\langle 1|
\nonumber \\
&\quad
+
G_{eg}
\left(
|0,1\rangle_{ab}\langle 1,0|
+
h.c
\right),
\end{align}
where $G_{eg}=\frac{1}{2}
(\lambda^a_{0,1} \beta^b_{0} + \lambda^b_{0,1} \beta^a_{0})$.
Eq.~\eqref{Eq:dissipative dynamics wc} and \eqref{Eq:H_D_TT} are valid only when the initial state of the transmons is restricted to the $\{|g\rangle, |e\rangle\}$ manifold.
The entanglement measure for all three cases is shown in Fig.~\ref{fig:fig:density evolution with and without cav}(d-f). All cases yield identical results. The entanglement measure used here is the logarithmic negativity $E_N(\hat\rho)$~\cite{PhysRevA.65.032314,Phys.Rev.Lett.95.090503}, defined as
\begin{equation}
E_N(\hat\rho) = \log_2 \left\| \hat\rho^{T_B} \right\|_1,
\end{equation}
where $\rho^{T_B}$ denotes the partial transpose of the reduced transmon-transmon density matrix with respect to one subsystem, and $\|\cdot\|_1$ is the trace norm.

Note that the results obtained in Fig.~\ref{fig:fig:density evolution with and without cav} for the $|e\rangle$ and $|g\rangle$ manifolds of the transmon can also be achieved for higher-level manifolds, provided that the energy gaps of those levels in the two transmons are in resonance with each other and detuned from the cavity.
%%%%%%%%%%%%%%%%%%%%%%%%%%%%%%%%%%%%%%%%%%%%%%%%%%%%%%%%%%

%%%%%%%%%%%%%%%%%%%%%%%%%%%%%%%%%%%%%%%%%%
\section{Continuous measurement of the two transmon}
%%%%%%%%%%%%%%%%%%%%%%%%%%%%%%%%%%%%%%%%%%%

We observe that, under the assumption that the cavity remains in its ground state, the coupled transmon–cavity–transmon system can be effectively reduced to a transmon–transmon system governed by Eq.~\eqref{Eq:dissipative dynamics wc} and Eq.~\eqref{Eq:H_D_TT}. In this effective description, dissipation arises solely from the spontaneous decay of each transmon from $|e\rangle$ to $|g\rangle$ into its respective bath.
The dissipative dynamics described by Eq.~\eqref{Eq:dissipative dynamics wc} corresponds to the unmonitored case, where photons emitted during spontaneous emission are not detected. As a result, the information carried away by the emitted photons is traced out, leading to a mixed-state evolution of the transmons.
In contrast, if the emitted photons are continuously monitored, the system evolves conditionally in a pure state, following quantum trajectories.
The model for continuous monitoring of the transmons is illustrated in Fig.~\ref{fig:System setup}(c). As shown in the figure, each transmon spontaneously emits photons into its respective environment, and the emitted photons are guided to corresponding detectors. The detectors measure these photons, or equivalently, the state of the environment.
Since the system and the environment are continuously entangled, such measurements project the state of the transmons conditioned on the state of the environment.
Note that in the effective transmon-transmon coupled system, the cavity remains in the ground state, and the transmons interact via virtual excitation of cavity photons. Therefore, monitoring of the cavity is not included in our continuous measurement setup. Since the Purcell decay can be neglected, the photons measured at the detectors are assumed to arise purely from spontaneous decay of the transmons.

To evaluate the trajectories of the monitored transmon–transmon system, we assume that initially there is no interaction between the system and the environment. The combined system–environment state is taken as
\begin{equation}
|\Psi(0)\rangle = |\psi(0)\rangle_a \otimes |\psi(0)\rangle_b,
\end{equation}
where
\begin{equation}
|\psi(0)\rangle_x = \left( c^x_g |g_x\rangle + c^x_e |e_x\rangle \right) \otimes |E^x_0\rangle,
\end{equation}
represents a product state of transmon $x$ and its corresponding environment ($x \in \{a,b\}$). Here, $|E^x_0\rangle$ denotes the initial state of the environment, and $|c^x_g|^2$ and $|c^x_e|^2$ are the probabilities of finding the transmon in $|g_x\rangle$ and $|e_x\rangle$, respectively.
As the system interacts with the environment, it undergoes spontaneous emission from $|e_x\rangle$ to $|g_x\rangle$. The joint system–environment state evolves as
\begin{equation}
|\Psi(t)\rangle = \hat{U}_{t,0} |\Psi(0)\rangle,
\end{equation}
where $U_{t,0}$ is the unitary time-evolution operator.
Instead of specifying the exact form of $\hat{U}_{t,0}$ and evaluating the state at some time t, we phenomenologically describe the infinitesimal time evolution as
\begin{equation}
\label{psi_t+dt}
|\Psi(t+dt)\rangle = |\psi(t+dt)\rangle_a \otimes |\psi(t+dt)\rangle_b,
\end{equation}
with
\begin{align}
\label{psi_t+dt_x}
|\psi(t + dt)\rangle_x =\; & c^x_e(t)\sqrt{p^x_g}\, |g_x\rangle |E^x_{1}\rangle \nonumber \\
& + \left( c^x_g(t) |g_x\rangle + c^x_e(t)\sqrt{1 - p^x_g}\, |e_x\rangle \right) |E^x_{0}\rangle .
\end{align}
Here, $|E^x_0\rangle$ corresponds to the environment state when no photon is emitted, while $|E^x_{1}\rangle$ corresponds to the state of the environment after emission of a photon due to the transition $|e_x\rangle \to |g_x\rangle$. 
The probability of emission in the interval $dt$ is $|c^x_e(t)|^2\, p^x_g$, where $p^x_g = \Gamma_{x,10}\, dt$, and $\Gamma_{x,10} = 1/T_{1,x}$ is the spontaneous emission rate of transmon $x$.
In Eq.~\eqref{psi_t+dt}, a product state is assumed due to the fact that the environments and detectors of each transmon are different and independent. Hence, each transmon becomes entangled with its respective environment, as given in Eq.~\eqref{psi_t+dt_x}.
The product state is modified at every time step $dt$ under the unitary evolution generated by the transmon-transmon interaction Hamiltonian $\hat{H}^D_{TT}$.

We consider continuous monitoring of the photons emitted into the environment from the transmons via photodetection. The detectors register a ``click'' when a photon is detected and ``no click'' otherwise.
When no photon is detected in either detector $a$ or $b$, corresponding to transmons $a$ and $b$, the state of the transmon–transmon system is projected as
\begin{equation}
|\Psi(t+dt)\rangle_s 
= \langle E^a_0 E^b_0 | \Psi(t+dt)\rangle 
= \hat{K}_{00} |\psi(t)\rangle_s,
\end{equation}
where
\begin{equation}
|\psi(t)\rangle_s =
\left( c^a_g |g_a\rangle + c^a_e |e_a\rangle \right)
\otimes
\left( c^b_g |g_b\rangle + c^b_e |e_b\rangle \right).
\end{equation}
If both transmons emit photons, the state is projected as
\begin{equation}
|\Psi(t+dt)\rangle_s 
= \langle E^a_1 E^b_1 | \Psi(t+dt)\rangle 
= \hat{K}_{11} |\psi(t)\rangle_s.
\end{equation}
If only one of the transmons emits a photon, the state is projected as
\begin{align}
|\Psi(t+dt)\rangle_s 
&= \langle E^a_1 E^b_0 | \Psi(t+dt)\rangle 
= \hat{K}_{10} |\psi(t)\rangle_s, \\
|\Psi(t+dt)\rangle_s 
&= \langle E^a_0 E^b_1 | \Psi(t+dt)\rangle 
= \hat{K}_{01} |\psi(t)\rangle_s.
\end{align}
We assume that the detectors operate in the Markovian regime, such that $\Gamma_{x,10}\, dt \ll 1$ for $x \in \{a,b\}$. $\hat{K}_{ij}$ is the Krauss operator where $(i,j)\in (0,1)$.

In a realistic experiment, photon detection is not perfect due to losses in the transmission channel or detector inefficiencies. As a result, some emitted photons remain undetected. To model this, we introduce a detection efficiency $\eta_x$ for each transmon $x \in \{a,b\}$.
The environment associated with each transmon is effectively divided into two channels: a monitored channel (detected photons) and an unmonitored channel (lost photons). 
The corresponding Kraus operators for this inefficient measurement process can be constructed from the operator
\begin{equation}
\hat{M} = \hat{M}_a \otimes \hat{M}_b,
\end{equation}
where
\begin{equation}
\hat{M}_x =
\begin{bmatrix}
1 & \sqrt{(1 - \eta_x)\Gamma_{x,10} dt}\,\hat{a}_{xL}^\dagger 
    + \sqrt{\eta_x \Gamma_{x,10} dt}\,\hat{a}_x^\dagger \\[8pt]
0 & \sqrt{1 - \Gamma_{x,10} dt}
\end{bmatrix}.
\label{eq:Kraus_matrix_x}
\end{equation}
The Kraus operators for inefficient detection, corresponding to no-click events in the monitored channels with possible loss, are given by:
\begin{itemize}
\item \text{No photon loss}
\begin{equation}
\hat{K}_{00:00} =
\langle E^a_0 E^{aL}_0 : E^b_0 E^{bL}_0 \,|\, \hat{M} \,|\, E^a_0 E^{aL}_0: E^b_0 E^{bL}_0 \rangle.
\end{equation}
\item \text{One photon lost from transmon $a$}
\begin{equation}
\hat{K}_{01:00} =
\langle E^a_0 E^{aL}_1 :E^b_0 E^{bL}_0 \,|\, \hat{M} \,|\, E^a_0 E^{aL}_0 :E^b_0 E^{bL}_0 \rangle.
\end{equation}
\item \text{One photon lost from transmon $b$}
\begin{equation}
\hat{K}_{00:01} =
\langle E^a_0 E^{aL}_0 :E^b_0 E^{bL}_1 \,|\, \hat{M} \,|\, E^a_0 E^{aL}_0: E^b_0 E^{bL}_0 \rangle.
\end{equation}
\item \text{Two photons lost (one from each transmon)}
\begin{equation}
\hat{K}_{01:01} =
\langle E^a_0 E^{aL}_1: E^b_0 E^{bL}_1 \,|\, \hat{M} \,|\, E^a_0 E^{aL}_0 :E^b_0 E^{bL}_0 \rangle.
\end{equation}
\end{itemize}
Here, $\hat{K}_{ij:kl} \equiv \hat{K}_{E^a_i E^{aL}_j : E^b_k E^{bL}_l}$, with $i,j,k,l \in \{0,1\}$. The states $|E^x_0\rangle$ and $|E^x_1\rangle$ correspond to no emission and emission into the monitored channel, respectively, while $|E^{xL}_0\rangle$ and $|E^{xL}_1\rangle$ correspond to no loss and emission into the unmonitored (loss) channel.
The environment creation operators act as
\(
\hat{a}_x^\dagger |E^x_0\rangle = |E^x_1\rangle, 
\quad
\hat{a}_{xL}^\dagger |E^{xL}_0\rangle = |E^{xL}_1\rangle,
\)
with all other actions yielding zero.%

The Kraus operators for inefficient detection, corresponding to click events in the transmon $a$ monitored channels with possible loss, are given by:
\begin{itemize}
\item \text{No photon loss}
\begin{equation}
\hat{K}_{10:00} =
\langle E^a_1 E^{aL}_0: E^b_0 E^{bL}_0 \,|\, \hat{M} \,|\, E^a_0 E^{aL}_0 :E^b_0 E^{bL}_0 \rangle.
\end{equation}
\item \text{One photon lost from transmon $b$}
\begin{equation}
\hat{K}_{10:01} =
\langle E^a_1 E^{aL}_0 :E^b_0 E^{bL}_1 \,|\, \hat{M} \,|\, E^a_0 E^{aL}_0 :E^b_0 E^{bL}_0 \rangle.
\end{equation}
\end{itemize}
Similarly, the Kraus operators for inefficient detection, corresponding to click events in the transmon $b$ monitored channels with possible loss, are given by:
\begin{itemize}
\item \text{No photon loss}
\begin{equation}
\hat{K}_{00:10} =
\langle E^a_0 E^{aL}_0: E^b_1 E^{bL}_0 \,|\, \hat{M} \,|\, E^a_0 E^{aL}_0 :E^b_0 E^{bL}_0 \rangle.
\end{equation}
\item \text{One photon lost from transmon $a$}
\begin{equation}
\hat{K}_{01:10} =
\langle E^a_0 E^{aL}_1 :E^b_1 E^{bL}_0 \,|\, \hat{M} \,|\, E^a_0 E^{aL}_0 :E^b_0 E^{bL}_0 \rangle.
\end{equation}
\end{itemize}
When both detectors corresponding to transmons $a$ and $b$ register clicks, the associated Kraus operator is given by
\begin{equation}
\hat{K}_{10:10} =
\langle E^a_1 E^{aL}_0: E^b_1 E^{bL}_0 \,|\, \hat{M} \,|\, E^a_0 E^{aL}_0 :E^b_0 E^{bL}_0 \rangle.
\end{equation}

The state update of the two-transmon system corresponding to the no-click event in the monitored channels (with possible loss into unmonitored channels) is given by
\begin{equation}
\begin{aligned}
\hat{\rho}_{\mathrm{tt}}(t + dt) = \frac{\hat{U}}{N} \Big(&
\hat{K}_{00:00}\, \hat{\rho}_{\mathrm{tt}}(t)\, \hat{K}_{00:00}^\dagger \\
&+ \hat{K}_{01:00}\, \hat{\rho}_{\mathrm{tt}}(t)\, \hat{K}_{01:00}^\dagger \\&+ \hat{K}_{00:01}\, \hat{\rho}_{\mathrm{tt}}(t)\, \hat{K}_{00:01}^\dagger \\
&+ \hat{K}_{01:01}\, \hat{\rho}_{\mathrm{tt}}(t)\, \hat{K}_{01:01}^\dagger
\Big)\hat{U}^\dagger,
\label{eq:K00}
\end{aligned}
\end{equation}
where $\hat{U} = \exp(-i\, \hat{H}^D_{TT}\, dt)$ is the unitary operator corresponding to the two-transmon interaction. This operation entangles the two transmon states. The normalization constant $N$ is given by
\begin{equation}
\begin{aligned}
N = \mathrm{Tr} \Big(&
\hat{K}_{00:00}\, \hat{\rho}_{\mathrm{tt}}(t)\, \hat{K}_{00:00}^\dagger + \hat{K}_{01:00}\, \hat{\rho}_{\mathrm{tt}}(t)\, \hat{K}_{01:00}^\dagger\\&
+ \hat{K}_{00:01}\, \hat{\rho}_{\mathrm{tt}}(t)\, \hat{K}_{00:01}^\dagger + \hat{K}_{01:01}\, \hat{\rho}_{\mathrm{tt}}(t)\, \hat{K}_{01:01}^\dagger
\Big).
\end{aligned}
\end{equation}
Here, photons emitted into the loss channels are inaccessible. Since the measurement outcomes of these unmonitored channels are inaccessible, their effects are incorporated by tracing over the corresponding state update. This results in a sum over all possible loss-channel outcomes in the state update. When the detection efficiency of the monitored channel is less than unity, the system evolves into a mixed state.
%%%%%%%%%%%%%%%%%%%%%%%%%%%%%%%%%%%%%%%%%%%%%%%%%%%%%%%%%%
\begin{figure*}
\centering
% -------- Row 1 --------
\includegraphics[width=0.3\linewidth]{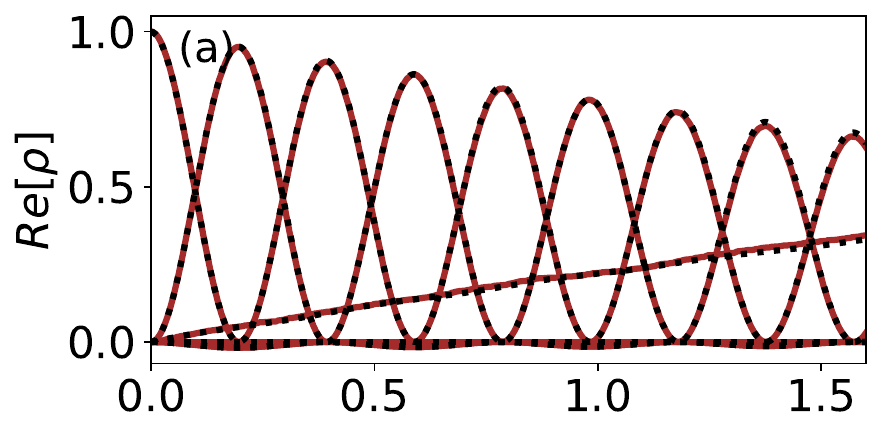}
\includegraphics[width=0.3\linewidth]{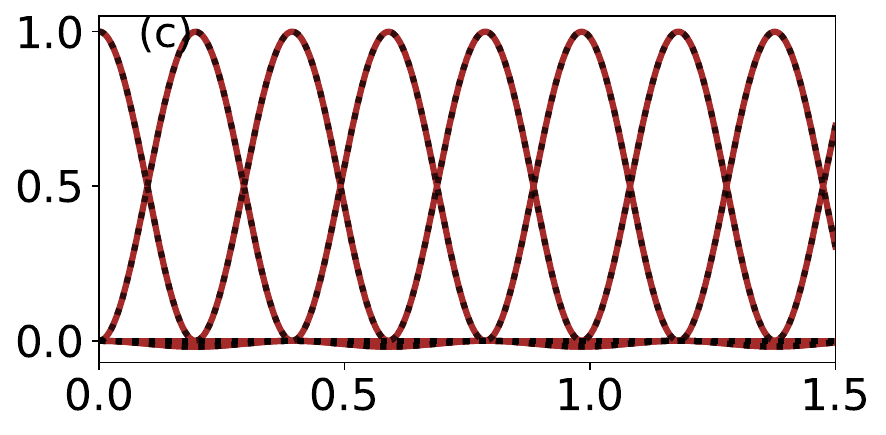}
\includegraphics[width=0.3\linewidth]{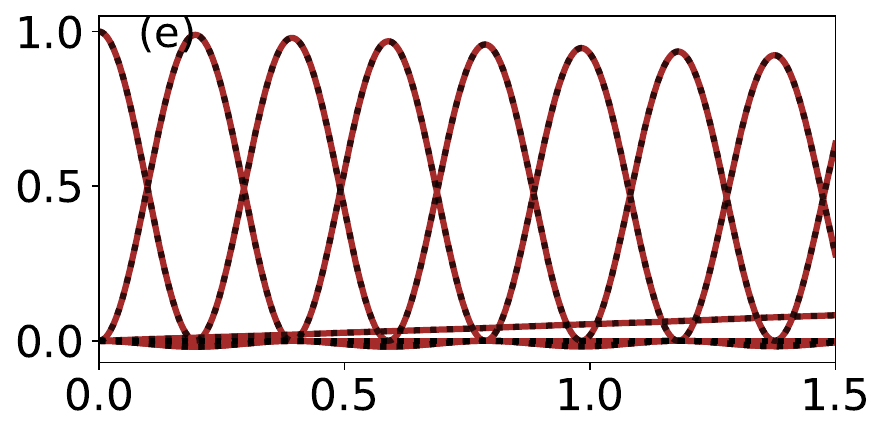}
\vspace{0.3cm}
% -------- Row 2 --------
\includegraphics[width=0.3\linewidth]{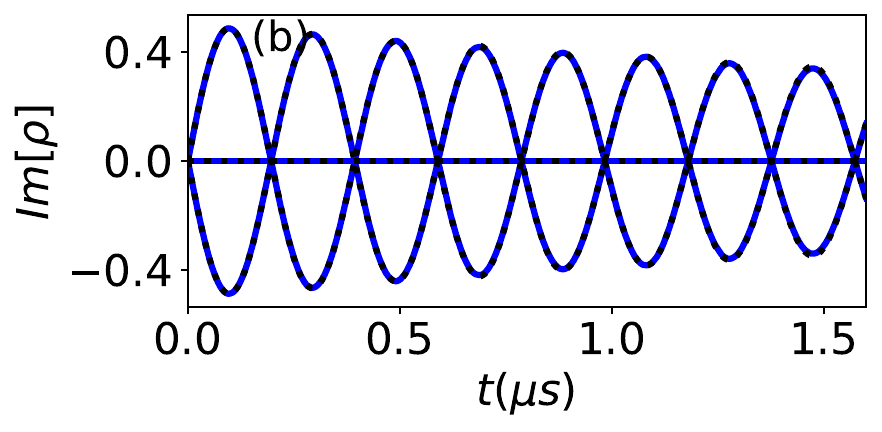}
\includegraphics[width=0.3\linewidth]{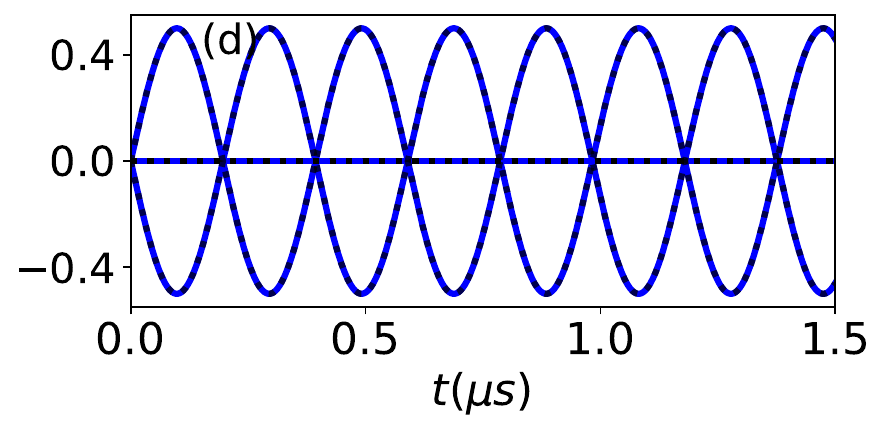}
\includegraphics[width=0.3\linewidth]{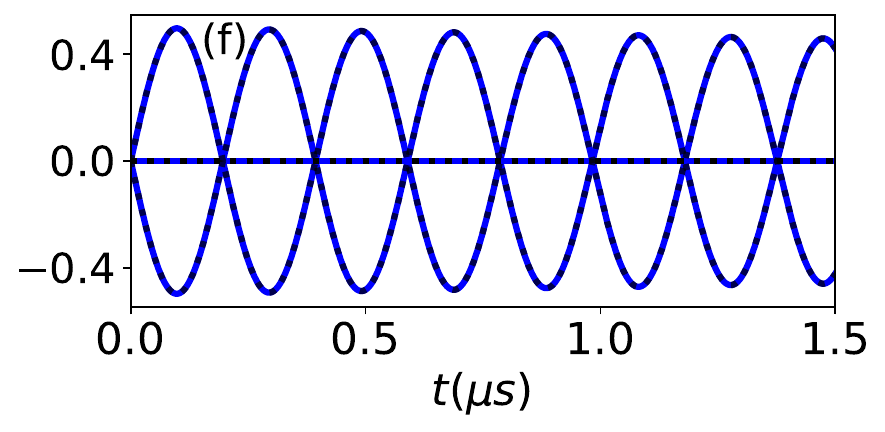}
\caption{Comparison between the trajectory ensemble average (dotted curve) and the average dynamics obtained from the master equation (solid curve). The top row shows the real part of the density matrix, while the bottom row shows the imaginary part.
(a,b) Unmonitored case with $\eta_x = 0$. The trajectory average is taken over 2000 realizations. The two approaches show excellent agreement.
(c,d) Perfect postselection case with $\eta_x = 1$. Since quantum jumps are postselected, all trajectories are identical and coincide with the master equation evolution conditioned on postselection.
(e,f) Imperfect postselection with $\eta_x = 0.8$. Here too, all trajectories remain identical and closely follow the corresponding master equation evolution. However, the coherence degrades due to leakage into the ground state arising from imperfect postselection.
The transmon system parameters are the same as in Fig.~\ref{fig:transmon_spec}. The decay rates are $\Gamma_{a,10} = 0.3~\mathrm{MHz}$ and $\Gamma_{b,10} = 0.2~\mathrm{MHz}$.}
\label{fig:density evolution compare traj and liouv}
\end{figure*}
%%%%%%%%%%%%%%%%%%%%%%%%%%%%%%%%%%%%%%%%%%%%%%%%%%%%%%%%%%%%
%
The state update when the transmon detectors registers a click (with possible loss into unmonitored channels) is given as follows.
\noindent\text{Click in detector $a$:}
\begin{equation}
\begin{aligned}
\hat{\rho}_{\mathrm{tt}}(t + dt) = \frac{\hat{U}}{N_a} \Big(&
\hat{K}_{10:00}\, \hat{\rho}_{\mathrm{tt}}(t)\, \hat{K}_{10:00}^\dagger \\
&+ \hat{K}_{10:01}\, \hat{\rho}_{\mathrm{tt}}(t)\, \hat{K}_{10:01}^\dagger
\Big)~\hat{U}^\dagger,
\label{eq:Ka}
\end{aligned}
\end{equation}
with normalization
\begin{equation}
\begin{aligned}
N_a = \mathrm{Tr} \Big(&
\hat{K}_{10:00}\, \hat{\rho}_{\mathrm{tt}}(t)\, \hat{K}_{10:00}^\dagger + \hat{K}_{10:01}\, \hat{\rho}_{\mathrm{tt}}(t)\, \hat{K}_{10:01}^\dagger
\Big).
\end{aligned}
\end{equation}
\noindent\text{Click in detector $b$:}
\begin{equation}
\begin{aligned}
\hat{\rho}_{\mathrm{tt}}(t + dt) = \frac{\hat{U}}{N_b} \Big(&
\hat{K}_{00:10}\, \hat{\rho}_{\mathrm{tt}}(t)\, \hat{K}_{00:10}^\dagger \\&+ \hat{K}_{01:10}\, \hat{\rho}_{\mathrm{tt}}(t)\, \hat{K}_{01:10}^\dagger
\Big)~\hat{U}^\dagger,
\label{eq:Kb}
\end{aligned}
\end{equation}
with normalization
\begin{equation}
\begin{aligned}
N_b = \mathrm{Tr} \Big(&
\hat{K}_{00:10}\, \hat{\rho}_{\mathrm{tt}}(t)\, \hat{K}_{00:10}^\dagger + \hat{K}_{01:10}\, \hat{\rho}_{\mathrm{tt}}(t)\, \hat{K}_{01:10}^\dagger
\Big).
\end{aligned}
\end{equation}
\noindent\text{Clicks in both detectors:}
\begin{equation}
\hat{\rho}_{\mathrm{tt}}(t + dt) =
\frac{\hat{U}~
\hat{K}_{10:10}\, \hat{\rho}_{\mathrm{tt}}(t)\, \hat{K}_{10:10}^\dagger
~\hat{U}^\dagger}{
\mathrm{Tr}\!\left(
\hat{K}_{10:10}\, \hat{\rho}_{\mathrm{tt}}(t)\, \hat{K}_{10:10}^\dagger
\right)
}.
\label{eq:Kab}
\end{equation}

The detectors continuously monitor the baths of the two transmons, and at each time interval $dt$, the state of the two-transmon system is conditionally updated based on the measurement outcomes, as described above. 
This monitoring process continues up to a time $t$, and the sequence of conditional state updates defines a single quantum trajectory.
By generating many such trajectories and taking their ensemble average, one recovers the unmonitored dynamics of the system, as obtained from solving Eq.~\eqref{Eq:dissipative dynamics wc}.
A comparison between the trajectory-averaged dynamics and the master equation evolution is shown in Fig.~\ref{fig:density evolution compare traj and liouv}(a-b), demonstrating excellent agreement.

\begin{figure}
\centering
\includegraphics[width=60mm]{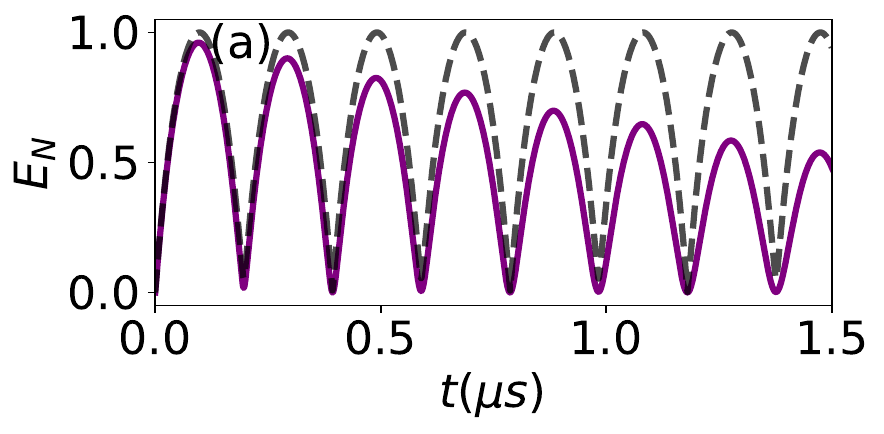}
\includegraphics[width=60mm]{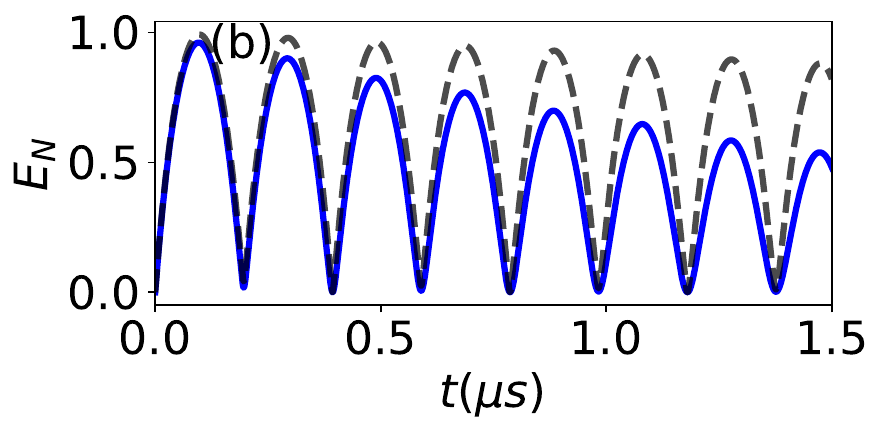}
\caption{(a) Comparison of the two-transmon entanglement measure between the unmonitored case $\eta_x = 0$ (solid) and the efficient postselection case $\eta_x = 1$ (dashed). While the entanglement in the unmonitored case decays, the entanglement under perfect postselection persists.
(b) Comparison of the two-transmon entanglement measure between the unmonitored case $\eta_x = 0$ (solid) and the inefficient postselection case $\eta_x = 0.8$ (dashed). While the entanglement in the unmonitored case decays, the entanglement under imperfect postselection degrades at a slower rate.} 
\label{fig:Postselect entanglement}
\end{figure}

Instead of averaging over all trajectories, one can postselect only those trajectories in which no quantum jump from $\ket{e}$ to $\ket{g}$ occurs, i.e., trajectories for which no detector click is registered up to a time $t$. Averaging over this subset of trajectories yields the postselected dynamics, as shown in Fig.~\ref{fig:density evolution compare traj and liouv}(c--f) for different detector efficiencies.
As observed, due to the absence of quantum jumps in the postselected ensemble, the coherence between the $\ket{e}$ and $\ket{g}$ states of the two transmons is preserved without decay. Consequently, the entanglement is maintained for a longer duration, as illustrated in Fig.~\ref{fig:Postselect entanglement}(a).
In this postselection scenario, the decay of entanglement is primarily limited by the detection efficiency, since missed detection events (due to inefficiency) effectively introduce unobserved jumps. This behavior is shown in Fig.~\ref{fig:Postselect entanglement}.
%%%%%%%%%%%%%%%%%%%%%%%%%%%%%%%%%%%%%%%%%%%%%%%%%%%%%%%%%%%%%%
\begin{figure*}
\centering
\includegraphics[width=180mm]{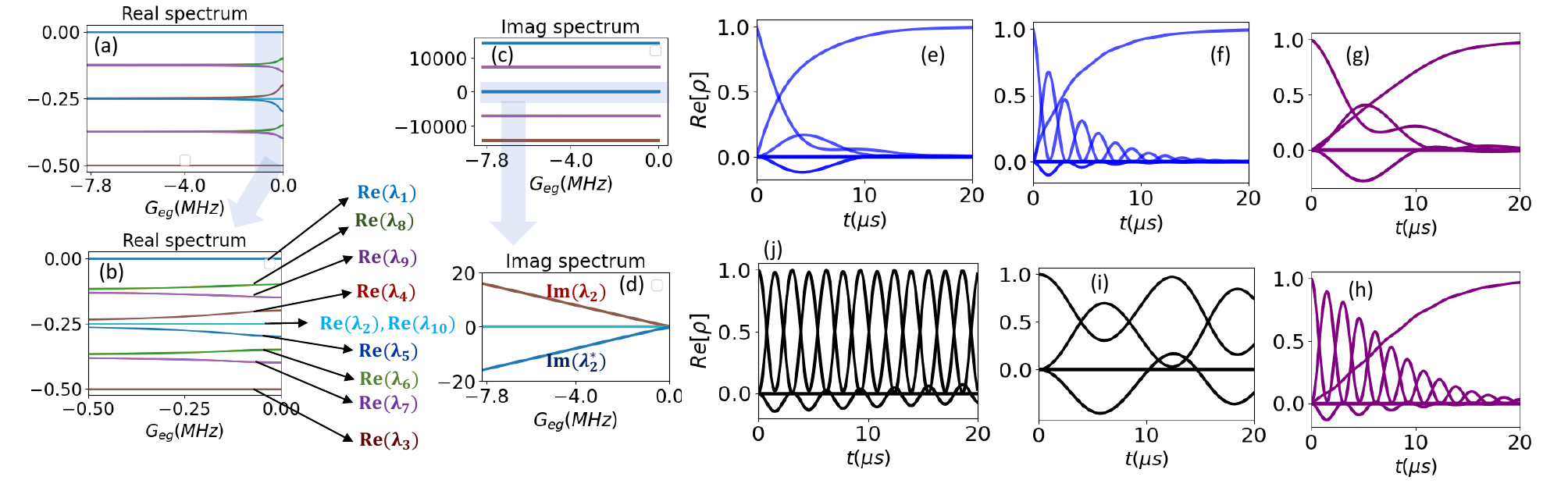}
\caption{Eigenvalue spectrum of the Liouvillian superoperator (a-d) and the corresponding evolution of density matrix elements (e-j). The real part of the spectrum is shown in (a), and a zoomed-in view near $G_{eg} = 0$, highlighted by the shaded region, is shown in (b). There are 16 eigenvalues in total, of which ten real parts are shown explicitly in (b). The remaining six arise as complex conjugate pairs of $\lambda_i$ with $i = 2,6,7,8,9,10$. The imaginary part of the spectrum is shown in (c). The contribution to the coherent dynamics arises from $\mathrm{Im}(\lambda_2)$ and $\mathrm{Im}(\lambda^*_2)$, as illustrated in (d).
Panels (e-j) show the evolution of the real part of density matrix elements in the unmonitored case ($\eta_x = 0$) for different values of $G_{eg}$. In (e,g,i), $G_{eg} = 0.2~\mathrm{MHz}$, while in (f,h,j), $G_{eg} = 1~\mathrm{MHz}$. The system is initialized in the state $\ket{e_a,g_b}$ and undergoes oscillations with $\ket{g_a,e_b}$. However, due to dissipation and the initial condition $C_1(0) = 1$, the ground state $\ket{g_a,g_b}$ becomes populated rapidly. As $G_{eg}$ increases, the oscillations become more pronounced, as seen in (f).
A similar behavior is observed in (g,h) for the imperfect postselection case ($\eta_x = 0.8$). In this case, $C_1(0) = 0.2$, leading to more visible oscillations and a slower decay compared to the unmonitored case. Panels (i,j) correspond to the perfect postselection case ($\eta_x = 1$), where $C_1(0) = 0$. Here, dissipation is strongly suppressed, resulting in sustained and more prominent oscillations.
All other parameters are the same as in Fig.~\ref{fig:density evolution compare traj and liouv}.}

\label{fig:Spectrum_eigenvalue_density}
\end{figure*}
%%%%%%%%%%%%%%%%%%%%%%%%%%%%%%%%%%%%555555

The state updates given in Eqs.~\eqref{eq:K00}--\eqref{eq:Kab} can be mapped to the following stochastic master equation:
\begin{align}
\frac{d\hat\rho_{tt}}{dt}
&= -i[\hat{H}^D_{TT},\hat\rho_{tt}] \nonumber \\[6pt]
&\quad + \sum_{x=a,b}\eta_x \Gamma_{x,10} \Big(
\langle 1|\hat\rho_{tt} |1\rangle_x\, \hat\rho_{tt}
- \frac{1}{2}\{ |1\rangle\langle 1|, \hat\rho_{tt} \}_x
\Big) \nonumber \\[6pt]
&\quad +\sum_{x=a,b} (1-\eta_x)\Gamma_{x,10}\, \mathcal{D}[|0\rangle\langle 1|]_x \hat\rho_{tt}
\nonumber\\
&\quad + \sum_{x=a,b}\left( \frac{\ket{0}_x\bra{1}\hat\rho_{tt} \ket{1}_x\bra{0}}
{\braket{\ket{1}_x\bra{1}}} - \hat\rho_{tt}  \right) dN_x.
\label{Eq:Postselect_master_eqn}
\end{align}
Here, $dN_x$ represents the counting process associated with the detector corresponding to transmon $x$, which takes the value 1 if a photon is detected in the time interval $dt$, and 0 otherwise. The mean value of this counting process is given by
\(
\langle dN_x \rangle = \Gamma_x\, \eta_x \, \langle e|\rho|e\rangle_x\, dt.
\)
Since photons emitted into the loss channels are not detected, the corresponding measurement records are inaccessible and effectively averaged out.
In the case of postselection, only those trajectories for which no detection events occur, i.e., $dN_x = 0$, are retained in the ensemble. Consequently, the master equation describing the postselected dynamics is obtained from Eq.~\eqref{Eq:Postselect_master_eqn} by omitting the jump term (last term).
The postselected dynamics obtained from this master equation is compared with the trajectory-averaged evolution in Fig.~\ref{fig:density evolution compare traj and liouv}(c-f) for different values of $\eta_x$. As observed, the two approaches are in excellent agreement.

Here, we stress that the monitoring of spontaneous emission from the $|e\rangle$ to $|g\rangle$ transition, as discussed above, can be generalized to higher excited-state manifolds of the two transmons, thereby enabling the investigation of the role of postselection in entanglement within these manifolds.
%%%%%%%%%%%%%%%%%%%%%%%%%%%%%%%%%%%%%%%%%%%%%%%%%%%%%%%%%%%%%%%%%%%%%%%
\section{Liouvillian Spectrum and Dynamics Near Exceptional Points}
%%%%%%%%%%%%%%%%%%%%%%%%%%%%%%%%%%%%%%%%%%%%%%%%%%%%%%%%%%%%%%%%%%%%%%
The master equation for the postselected dynamics is obtained from Eq.~\eqref{Eq:Postselect_master_eqn} as
\begin{align}
\frac{d\hat{\rho}_{\mathrm{tt}}}{dt}
&= -i[\hat{H}^D_{TT},\hat{\rho}_{\mathrm{tt}}]
+ \sum_{x=a,b} (1-\eta_x)\Gamma_{x,10}\, \mathcal{D}[|0\rangle\langle 1|]_x \hat{\rho}_{\mathrm{tt}}\nonumber \\[6pt]
&\quad + \sum_{x=a,b}\eta_x \Gamma_{x,10}\Big(
\langle 1|\hat{\rho}_{\mathrm{tt}} |1\rangle_x\, \hat{\rho}_{\mathrm{tt}}
- \frac{1}{2}\{ |1\rangle\langle 1|, \hat{\rho}_{\mathrm{tt}} \}_x
\Big).
\label{Eq:Postselect_master_eqn_liouv}
\end{align}
At $\eta_x = 0$, this reduces to the unmonitored master equation given in Eq.~\eqref{Eq:dissipative dynamics wc}. The postselected master equation is nonlinear due to the trace-preserving term 
$\eta_x \Gamma_{x,10} \langle 1|\hat{\rho}_{\mathrm{tt}} |1\rangle_x\, \hat{\rho}_{\mathrm{tt}}$.
The dynamics generated by this nonlinear master equation can equivalently be obtained by first evolving the system under the corresponding linear (unnormalized) master equation—i.e., without the nonlinear term—and subsequently normalizing the density matrix by dividing by its trace, $\mathrm{Tr}(\hat{\rho}_{\mathrm{tt}})$.
In the absence of the nonlinear term, the evolution can be written in terms of a Liouvillian superoperator as
\begin{equation}
\label{Eq:Liouvillian ME}
    \frac{d\ket\rho_{\mathrm{tt}}}{dt} = \hat{\mathcal{L}}\ket\rho_{\mathrm{tt}}.
\end{equation}
Here, $\hat{\rho}_{\mathrm{tt}}$ is expressed in vectorized form as a $16 \times 1$ column vector, obtained via column-stacking of the density matrix, and $\hat{\mathcal{L}}$ is the corresponding $16 \times 16$ Liouvillian superoperator. The explicit form of $\hat{\mathcal{L}}$ is given in the Appendix.
The solution can be written in terms of the decomposition over the right eigenvectors of $\hat{\mathcal{L}}$:
\begin{equation}
    \label{decomposition}
    \ket{\rho(t)}_{tt}=\sum_{i=0}^{15} C_i(0)\, e^{-\lambda_i t}\ket{R}_i,
\end{equation}
where $\ket{R}_i$ is the right eigenvector and $\lambda_i$ is the corresponding eigenvalue. The coefficients are given by
\(
    C_i(0)=\langle L_i|\rho(0)_{tt}\rangle,
\)
$\langle L_i|$ is the left eigenvector.
The real and imaginary parts of the eigenvalue spectrum are shown in Fig.~\ref{fig:Spectrum_eigenvalue_density}(a-d). For the initial state $|1_a 0_b\rangle$, the eigenvalues $\lambda_1=0$, $\lambda_2$, $\lambda_2^*$, $\lambda_4$, and $\lambda_5$ determine the dynamics of the system. This is because, for the given initial state, $C_1(0)$, $C_2(0)$, $C_4(0)$, and $C_5(0)$ are nonzero, while all other coefficients vanish. The eigenvalues $\lambda_2$ and $\lambda_2^*$ have coefficients $C_2(0)$ and $C^*_2(0)$, respectively.
The eigenvalues are independent of the detection efficiency $\eta_x$, whereas the coefficients $C_i(0)$ depend on it. Furthermore, as the value of $\eta_x$ increases, $C_1(0)$ decreases and becomes zero at $\eta_x=1$.

The dynamics of the density matrix elements are governed by the differences between the contributing eigenvalues.
This arises due to the normalization by the trace of the density matrix to ensure trace preservation.
These differences increase as the magnitude of the coupling $G_{eg}$ decreases, as shown in Fig.~\ref{fig:Spectrum_eigenvalue_density}(b--d).
The decay of the dynamics is governed by the gap in the real parts of the eigenvalues, while the frequency of the coherent oscillations is determined by the gap in the imaginary parts of the corresponding eigenvalues. 
This is illustrated in Fig.~\ref{fig:Spectrum_eigenvalue_density}(e-j) for $\eta_x=0$, $\eta_x=0.8$, and $\eta_x=1$, and for two different values of $G_{eg}$: one with a large gap ($G_{eg}=0.2\,\text{MHz}$) and the other with a small gap ($G_{eg}=1\,\text{MHz}$).
A larger gap leads to faster decay with fewer oscillations, whereas a smaller gap results in slower decay with more pronounced oscillations. Compared to the case $\eta_x<1$, when $\eta_x=1$ there is no contribution from $\lambda_1=0$, since $C_1(0)=0$ at $\eta_x=1$, and the dynamics persist for a longer time for both small and large gaps. This is evident from a comparison of Fig.~\ref{fig:Spectrum_eigenvalue_density}(e--h) with Fig.~\ref{fig:Spectrum_eigenvalue_density}(i--j).

In the interaction frame of the two-transmon system, an exceptional point (EP) can be observed. In this frame, the Hamiltonian reduces to the interaction term $\hat{H}_I = G_{eg}\left( \ket{0,1}_{ab}\bra{1,0} + \text{h.c.} \right)$.
Substituting this Hamiltonian into Eq.~\eqref{Eq:Postselect_master_eqn_liouv}, we obtain the Liouvillian superoperator in the interaction frame, whose eigenvalue spectrum is shown in Fig.~\ref{fig:eigenvalue spectrum int pic}. The explicit expressions for the eigenvalues are provided in the Appendix \ref{appendix A}. The exceptional point occurs at $(\Gamma_{b,10}-\Gamma_{a,10})/4$ (indicated by the red dotted line in the figure).
For $\eta_x=1$, the dynamics are governed by the eigenvalues $\lambda_2$, $\lambda_4$, and $\lambda_5$, since the corresponding coefficients $C_i(0)$ are the only nonzero ones.
%%%%%%%%%%%%%%%%%%%%%%%%%%%%%%%%%%%%%%%%%%%%%%%%%%%%%%%%55
\begin{figure}
\centering
\includegraphics[width=70mm]{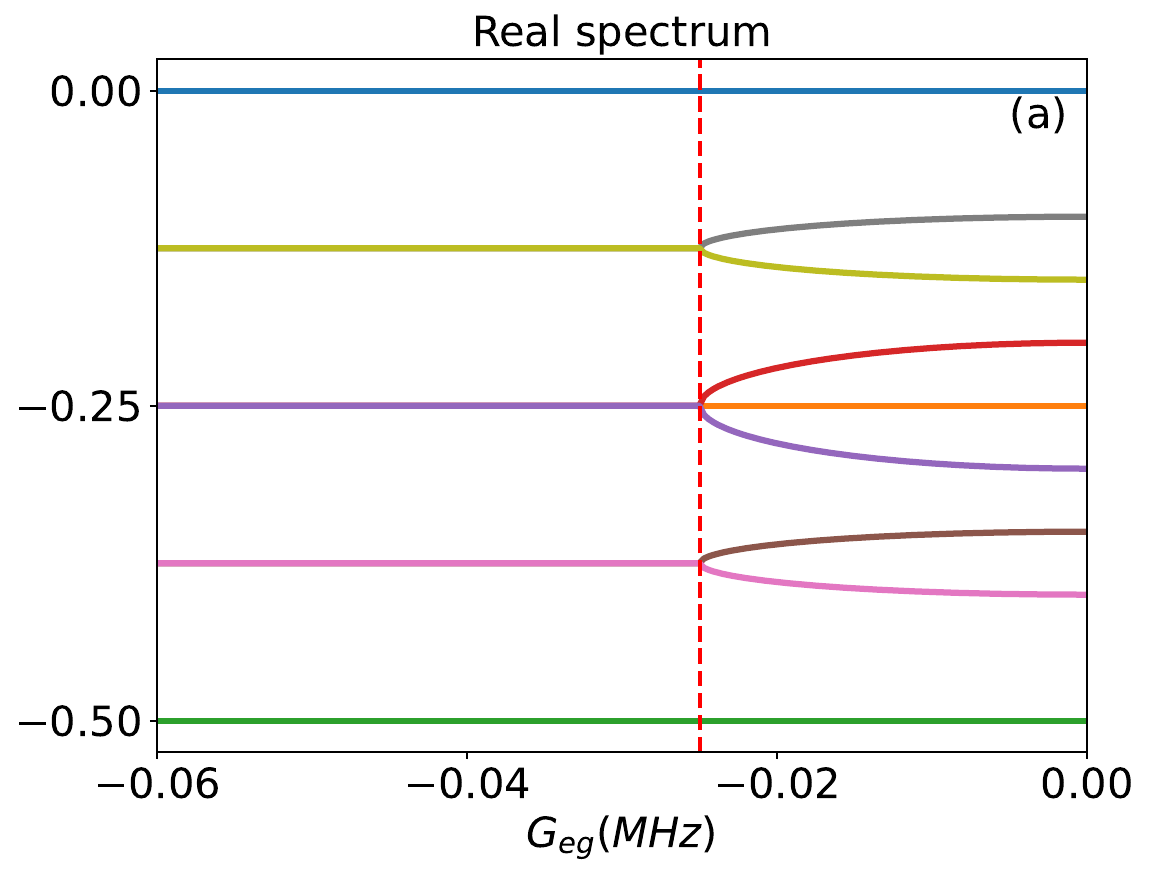}
\includegraphics[width=70mm]{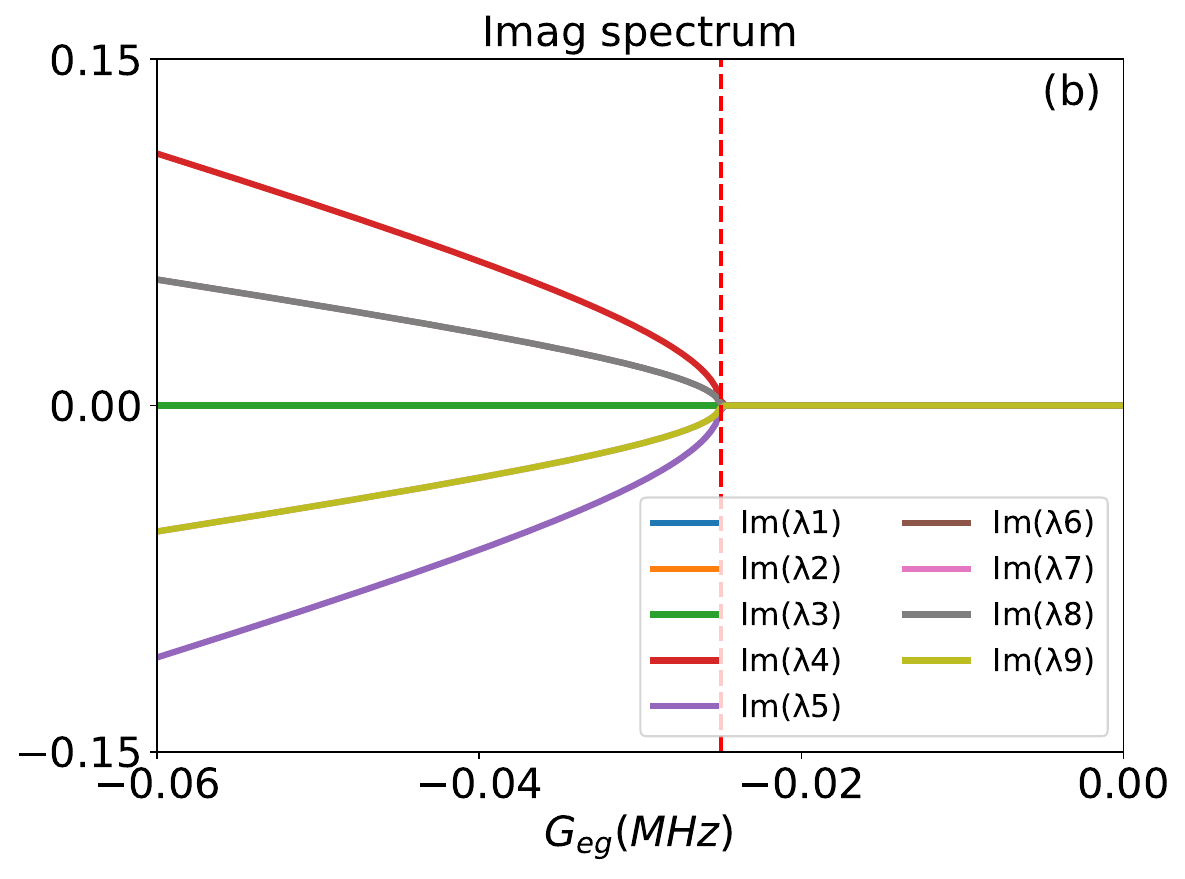}
\caption{Eigenvalue spectrum of the Liouvillian superoperator in the interaction frame. (a) Real part of the spectrum. (b) Imaginary part of the spectrum. The exceptional point (EP), located at $G_{eg} = 0.025$, is indicated by the red dotted line. The parameters are the same as in Fig.~\ref{fig:eigenvalue spectrum int pic}.}
\label{fig:eigenvalue spectrum int pic}
\end{figure}

\begin{figure}
\centering
\includegraphics[width=0.5\linewidth]{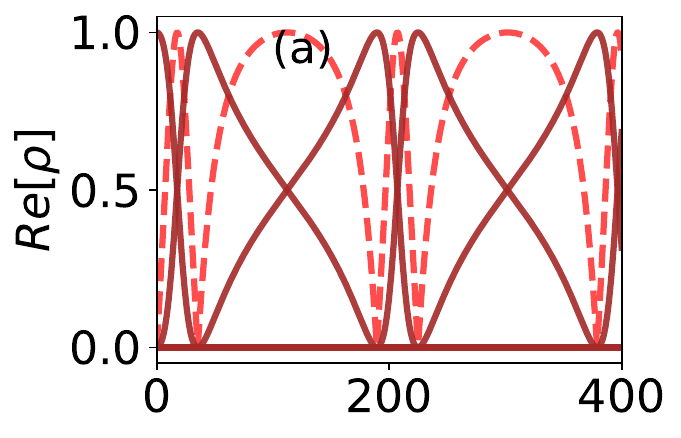}
\includegraphics[width=0.45\linewidth]{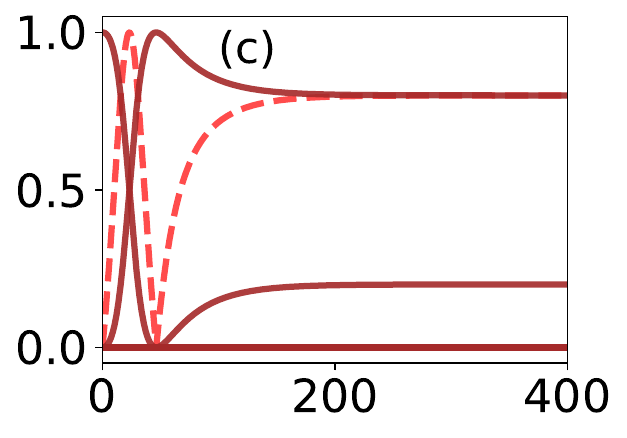}
\vspace{0.3cm}
\includegraphics[width=0.5\linewidth]{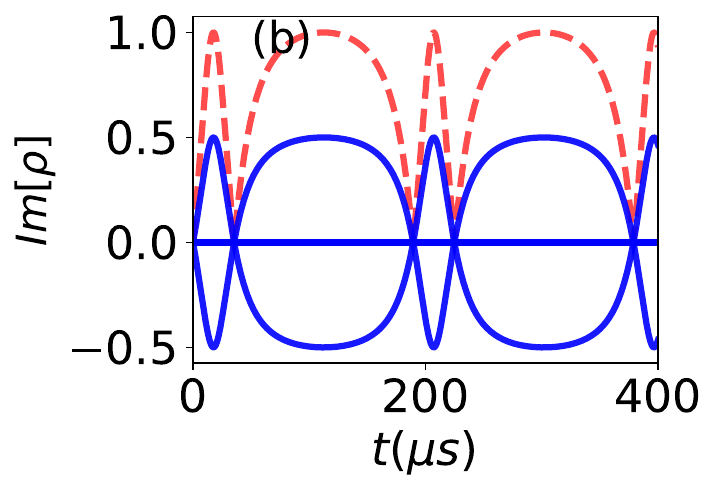}
\includegraphics[width=0.45\linewidth]{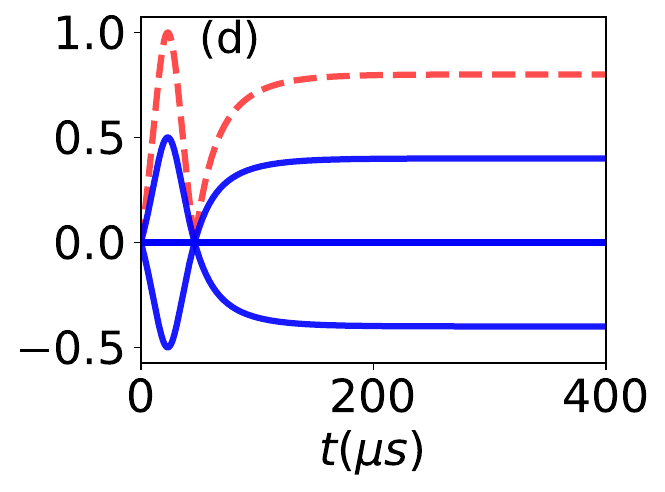}
\caption{Real (a,c) and imaginary (b,d) parts of the density matrix evolution. (a,b) Evolution for $G_{eg} = 0.03~\text{MHz}$, which lies to the left of the exceptional point (EP). This resembles the $\mathcal{PT}$-symmetric phase dynamics. (c,d) Evolution for $G_{eg} = 0.02~\text{MHz}$, which lies to the right of the EP. This resembles the broken $\mathcal{PT}$-symmetric phase dynamics. The entanglement dynamics in both the broken and unbroken $\mathcal{PT}$-symmetric phases are shown by the red dotted curve.}
\label{fig:int pic evolution}
\end{figure}
On the left side of the exceptional point (EP), the eigenvalues of the Liouvillian have both real and imaginary parts. The real parts of the contributing eigenvalues are equal. As a result, the evolution of the density matrix elements exhibits no decay due to normalization, and the system undergoes purely oscillatory dynamics, as shown in Fig.~\ref{fig:int pic evolution}(a-b). The entanglement also oscillates correspondingly without any decay. This behavior resembles the $\mathcal{PT}$-symmetric case, where the dynamics are purely oscillatory due to purely imaginary eigenvalues of the Liouvillian.
On the right side of the EP, the eigenvalues are purely real, and the contributing eigenvalues $\lambda_4$ and $\lambda_5$ take different values. This leads to an effective decay in the dynamics. Since there is no imaginary part in the eigenvalues, no oscillations are observed. The dynamics and the entanglement in this case are shown in Fig.~\ref{fig:int pic evolution}(c-d). This regime resembles the $\mathcal{PT}$-symmetry-broken case, where the dynamics are purely dissipative due to purely real eigenvalues of the Liouvillian.

%%%%%%%%%%%%%%%%%%%%%%%%%%%%%%%%%%%%%%%%%%%%%%%%%%%%%%%%%%
\section{Conclusion}
%%%%%%%%%%%%%%%%%%%%%%%%%%%%%%%%%%%%%%%%%%%%%%%%%%%%%%%%%%

The role of continuous measurement and postselection in the dynamics and entanglement of a transmon-cavity-transmon coupled system has been investigated. In the dispersive regime, the interaction between the two transmons is effectively mediated via virtual excitation of the cavity, resulting in an effective transmon--transmon coupling.
Continuous monitoring of the spontaneous emission from each transmon is performed via photodetection of the environment. By incorporating detector inefficiencies, both ideal and realistic measurement scenarios are analyzed. The trajectories that do not undergo quantum jumps are postselected, and the corresponding ensemble-averaged dynamics are obtained.
The results demonstrate that the postselected average dynamics significantly slow down the decay of entanglement compared to the unmonitored case, highlighting the important role of measurement backaction and information extraction in preserving quantum correlations.
The dynamics are further analyzed using the Liouvillian superoperator spectrum. In the interaction frame, the emergence of an exceptional point (EP) is identified, along with the associated broken and unbroken $\mathcal{PT}$-symmetric phases. These phases are shown to strongly influence both the system dynamics and the corresponding entanglement behavior.
We note that the results obtained in this work can be extended to higher excited-state manifolds of the transmons. Furthermore, while the transmons considered here are capacitively coupled via a static cavity, alternative tunable couplers can be employed to implement two-qubit gates~\cite{BliasRevModPhys.93.025005}.
Overall, our study shows how continuous measurement and postselection provide insight into controlling and preserving entanglement in dissipative quantum systems, with potential applications in quantum information processing and engineered non-Hermitian dynamics.
%%%%%%%%%%%%%___________________________

\section*{Acknowledgement}
This work is supported by MoE, Government of India (Grant No. MoE-STARS/STARS-2/2023-0161).
%%%%%%%%%%%%%%%%%%%%%%%%%%%%%%%%%%%%%%%%%%
\bibliography{ref}
%%%%%%%%%%%%%%%%%%%%%%%%%%%%%%%%%%%%%%%%%%%%%%%%%%%%%%%%%%%%5
\vspace{15mm}
\appendix
\section{Liouvillian superoperator}
\label{appendix A}

\begin{widetext}
The Liouvillian superoperator defined in Eq.~\eqref{Eq:Liouvillian ME} is given by
\begin{equation}
\label{liouvillian general}
\hat{\mathcal{L}} = \hat{\mathcal{L}}_{\mathrm{H}} + \hat{\mathcal{L}}_{\mathrm{diss}} + \hat{\mathcal{L}}_{\mathrm{jump}},
\end{equation}
where
\begin{align}
\hat{\mathcal{L}}_{\mathrm{H}} &= -i \left( \hat{H}^D_{TT} \otimes \hat{I}_4 - \hat{I}_4 \otimes \hat{H}^{D~(T)}_{TT} \right), \\[6pt]
\hat{\mathcal{L}}_{\mathrm{diss}} &= \sum_{x=a,b} -\frac{1}{2}\,\Gamma_{x,10}\,\eta_x 
\left( |e\rangle_x\langle e| \otimes \hat{I}_4 + \hat{I}_4 \otimes |e\rangle_x\langle e| \right), \\[6pt]
\hat{\mathcal{L}}_{\mathrm{jump}} &= \sum_{x=a,b} \Gamma_{x,10} (1-\eta_x) \left( 
|g\rangle_x\langle e| \otimes |g\rangle_x\langle e|
- \frac{1}{2} |e\rangle_x\langle e| \otimes \hat{I}_4
- \frac{1}{2} \hat{I}_4 \otimes |e\rangle_x\langle e|
\right).
\label{Liouvillian structure}
\end{align}
Here, $\hat{H}^{D~(T)}_{TT}$ is the transpose of $\hat{H}^{D}_{TT}$.
In the interaction frame of the Hamiltonian, the superoperator is given by
\begin{equation}
\small
\setlength{\arraycolsep}{3pt}
\hat{\mathcal{L}} =
\left(
\begin{array}{*{16}{c}}
0 & 0 & 0 & 0 & 0 & \Gamma_{\eta_b} & 0 & 0 & 0 & 0 & \Gamma_{\eta_a} & 0 & 0 & 0 & 0 & 0 \\
0 & B & iG_{eg} & 0 & 0 & 0 & 0 & 0 & 0 & 0 & 0 & \Gamma_{\eta_a} & 0 & 0 & 0 & 0 \\
0 & iG_{eg} & A & 0 & 0 & 0 & 0 & \Gamma_{\eta_b} & 0 & 0 & 0 & 0 & 0 & 0 & 0 & 0 \\
0 & 0 & 0 & A+B & 0 & 0 & 0 & 0 & 0 & 0 & 0 & 0 & 0 & 0 & 0 & 0 \\
0 & 0 & 0 & 0 & B & 0 & 0 & 0 & -iG_{eg} & 0 & 0 & 0 & 0 & 0 & \Gamma_{\eta_a} & 0 \\
0 & 0 & 0 & 0 & 0 & 2B & iG_{eg} & 0 & 0 & -iG_{eg} & 0 & 0 & 0 & 0 & 0 & \Gamma_{\eta_a}  \\
0 & 0 & 0 & 0 & 0 & iG_{eg} & A+B & 0 & 0 & 0 & -iG_{eg} & 0 & 0 & 0 & 0 & 0 \\
0 & 0 & 0 & 0 & 0 & 0 & 0 & A+2B & 0 & 0 & 0 & -iG_{eg} & 0 & 0 & 0 & 0 \\
0 & 0 & 0 & 0 & -iG_{eg} & 0 & 0 & 0 & A & 0 & 0 & 0 & 0 & \Gamma_{\eta_b} & 0 & 0 \\
0 & 0 & 0 & 0 & 0 & -iG_{eg} & 0 & 0 & 0 & A+B & iG_{eg} & 0 & 0 & 0 & 0 & 0 \\
0 & 0 & 0 & 0 & 0 & 0 & -iG_{eg} & 0 & 0 & iG_{eg} & 2A & 0 & 0 & 0 & 0 & \Gamma_{\eta_b} \\
0 & 0 & 0 & 0 & 0 & 0 & 0 & -iG_{eg} & 0 & 0 & 0 & 2A+B & 0 & 0 & 0 & 0 \\
0 & 0 & 0 & 0 & 0 & 0 & 0 & 0 & 0 & 0 & 0 & 0 & A+B & 0 & 0 & 0 \\
0 & 0 & 0 & 0 & 0 & 0 & 0 & 0 & 0 & 0 & 0 & 0 & 0 & A+2B & iG_{eg} & 0 \\
0 & 0 & 0 & 0 & 0 & 0 & 0 & 0 & 0 & 0 & 0 & 0 & 0 & iG_{eg} & 2A+B & 0 \\
0 & 0 & 0 & 0 & 0 & 0 & 0 & 0 & 0 & 0 & 0 & 0 & 0 & 0 & 0 & 2A+2B
\end{array}
\right)
\end{equation}

where $A=\Gamma_{a,10}/2$, $B=\Gamma_{b,10}/2$, and $\Gamma_{\eta_x}=(1-\eta_x)\,\Gamma_{x,10}$. The corresponding eigenvalues are given by
\begin{equation}
\begin{aligned}
\lambda_1 &= 0 \quad &&(\text{multiplicity } 1), \\[6pt]
\lambda_2 &= A + B \quad &&(\text{multiplicity } 4), \\[6pt]
\lambda_3 &= 2(A + B) \quad &&(\text{multiplicity } 1), \\[6pt]
\lambda_{4,5} &= A + B \;\pm\; \sqrt{(A - B)^2 - 4 G_{eg}^2} 
\quad &&(\text{each multiplicity } 1), \\[6pt]
\lambda_{6,7} &= \frac{3}{2}(A + B) \;\pm\; \frac{1}{2}\sqrt{(A - B)^2 - 4 G_{eg}^2} 
\quad &&(\text{each multiplicity } 2), \\[6pt]
\lambda_{8,9} &= \frac{1}{2}(A + B) \;\pm\; \frac{1}{2}\sqrt{(A - B)^2 - 4 G_{eg}^2} 
\quad &&(\text{each multiplicity } 2).
\end{aligned}
\end{equation}
Notably, the eigenvalues are independent of the measurement efficiency $\eta_x$. 
The Liouvillian corresponding to the lab frame is also independent of the measurement efficiency $\eta_x$, since the additional diagonal contributions arising from $\hat{H}^{D}_{\mathrm{TT}}$ adds only to the diagonal of the Liouvillian matrix.

\end{widetext}

\end{document}